\documentclass[twocolumn]{aastex61}
\usepackage{graphicx}
\usepackage{natbib}
\newcommand\aastex{AAS\TeX}

\received{August 17, 2018}
\revised{January 13, 2019}
\accepted{January 15, 2019}
\submitjournal{ApJ}

\shorttitle{\aastex\ Bar Fraction}
\shortauthors{Lee et al. 2019}

\begin{document}
\defcitealias{Ann15}{Ann15}
\defcitealias{RC3}{RC3}
\defcitealias{Lau02}{Lau02}

\title{Bar Fraction in Early- and Late-type Spirals}

\correspondingauthor{Myeong-Gu Park}
\email{mgp@knu.ac.kr}

%

\author[0000-0002-0786-7307]{Yun Hee Lee}
\affil{Department of Astronomy and Atmospheric Sciences, Kyungpook National University, Daegu, 41566, Korea} 

\author{Hong Bae Ann}
\affiliation{Dong-Pusan College, Busan, 48000, Korea}

\author{Myeong-Gu Park}
\affil{Department of Astronomy and Atmospheric Sciences, Kyungpook National University, Daegu, 41566, Korea} 
\affil{Research and Training Team for Future Creative Astrophysicists and Cosmologists (BK21 Plus Program), Kyungpook National University, Daegu, 41566, Korea} 







\begin{abstract}

Bar fractions depend on the properties of the host galaxies, which are closely related to the formation and evolution of bars. However, observational studies do not provide consistent results. We investigate the bar fraction and its dependence on the properties of the host galaxies by using three bar classification methods: visual inspection, ellipse fitting method, and Fourier analysis. Our volume-limited sample consists of 1,698 spiral galaxies brighter than $M_{\rm r}=-15.2$ with $z < 0.01$ from SDSS/DR7 visually classified by \citet{Ann15}. We first compare the consistency of classification among the three methods. Automatic classifications detect visually determined strongly barred galaxies with the concordance of $74\% \sim 85\%$. However, they have some difficulty in identifying bars, in particular, in bulge-dominated galaxies, which affects the distribution of bar fraction as a function of the Hubble type. We obtain, for the same sample, different bar fractions of 63\%, 48\%, and 36\% by visual inspection, ellipse fitting, and Fourier analysis, respectively. The difference is mainly due to how many weak bars are included. Moreover, we find the different dependence of bar fraction on the Hubble type for strong versus weak bars: SBs are preponderant in early-type spirals whereas SABs in late-type spirals. This causes a contradictory dependence on host galaxy properties when different classification methods are used. We propose that strong bars and weak bars experience different processes for their formation, growth, and dissipation by interacting with different inner galactic structures of early-type and late-type spirals. 

\end{abstract}




\section{Introduction} \label{chap1}

Early theoretical studies showed that stellar disks lacking random motions are generally unstable, and rapidly leads to the formation of bars \citep{1964ApJ...139.1217T, 1973ApJ...186..467O}. On the other hand, spherical components such as bulges and halos stabilize the stellar disks and delay the bar formation \citep{1973ApJ...186..467O, 1986MNRAS.221..213A}. However, once a bar forms, spherical components help the bar grow stronger \citep{1981A&A....96..164C, 2002MNRAS.330...35A, 2002ApJ...569L..83A} by depriving of the angular momentum and energy from the bars \citep{1970IAUS...38..318K, 1972MNRAS.157....1L, 1979ApJ...227..714K, 1980A&A....89..296S, 1984ApJ...282L...5T, 1985MNRAS.213..451W, 1991MNRAS.250..161L, 1992ApJ...400...80H, 1996ASPC...91..309A, 2002ApJ...569L..83A, 2003MNRAS.341.1179A, 1998ApJ...493L...5D, 2000ApJ...543..704D}.

Bars drive gas and stars into the galactic center by large-scale streaming motions \citep{1979ApJ...233...67R, 1981ApJ...246..740V, 1981ApJ...247...77S, 1984MNRAS.209...93S, 1983IAUS..100..215P, 1985A&A...150..327C, 1992MNRAS.259..345A, 1993RPPh...56..173S}. The bar-driven stars trapped in an elongated orbit induce the bar to become more eccentric \citep{1985A&A...150..327C, 2003MNRAS.341.1179A} and gas inflow driven by the bar builds up the central mass concentration (CMC) \citep{1990ApJ...363..391P, 2000JKAS...33....1A, 2002MNRAS.330...35A}. Subsequently, the growth of CMC broadens radial and vertical resonance regions, and induces slow secular evolution, which creates bulge-like structures such as pseudo-bulges and peanut-shaped bulges, and dissolves the bar itself \citep{1990ApJ...363..391P, 1990ApJ...361...69H, 1993ApJ...409...91H, 1993IAUS..153..209K, 1996ApJ...462..114N, 2006ApJ...645..209D}. These are the general picture for bar formation, evolution, and destruction that we have learned from model and simulation studies.

We expect to find some clues to the formation, growth, and destruction of bars from the observed bar fraction in different galaxies, which is the integrated outcome of the processes bars have experienced. Bars, including both strong and weak bars, have been found over 60\% of disk galaxies in the local universe \citep{RC3, 2010ApJS..190..147B, 2015ApJS..217...32B, Ann15}. The bar fraction has been widely known to depend strongly on the Hubble sequence, mass, color, and bulge prominence \citep{2008ApJ...675.1141S}. However, detailed observational tendencies are still controversial: bar fractions are reported to increase toward early-type spirals which are massive, red, gas-poor, and bulge-dominated \citep{2008ApJ...675.1141S, Agu09, 2009ApJ...692L..34L, 2013ApJ...779..162C, 2015A&A...580A.116G, 2016AA...595A..67C}, or increase toward the opposite direction of Hubble sequence, i.e., toward late-type spirals which are less massive, blue, gas-rich, and disk dominated \citep{2008ApJ...675.1194B, 2009A&A...497..713B, Agu09, 2009ApJ...696..411W, 2015ApJS..217...32B, 2015MNRAS.446.3749Y, 2018Erwin}. In addition, other studies have shown the bimodal distribution of the bar fraction, with each peak in early-type and in late-type spirals \citep{RC3, 1999ASPC..187...72K, 2000AJ....119..536E, 2010ApJ...714L.260N, 2011MNRAS.411.2026M, 2012ApJ...745..125L, 2012ApJS..198....4O, 2016A&A...587A.160D}. 

How can we understand this apparent disagreement? \citet{2010ApJ...714L.260N} have explained the reason for the discrepancy as the different mass range of sample galaxies each study had used, while \citet{2018Erwin} considered the angular resolution from the combination of the FWHM for telescope and the distance to sample galaxies. In this paper, we investigate the effect of classification methods to detect barred galaxies and the different bar definitions in the studies.

We suspect that different methods to identify bars may have contributed to the inconsistent results on the bar fraction as a function of Hubble sequence. In previous studies, the high bar fractions in early-type spirals were mainly obtained by visual inspection or Fourier analysis \citep{2008ApJ...675.1141S, 2013ApJ...779..162C, Agu09, 2009ApJ...692L..34L}, whereas the opposite results were derived mainly by the ellipse fitting method \citep{2008ApJ...675.1194B, 2009A&A...497..713B, Agu09}. Since the sample for the classification itself can greatly affect the outcome, we need to apply each bar detection method to the same sample, and compare the results. We chose visual inspection, ellipse fitting, and Fourier analysis methods in this paper.  

In addition, we can not ignore the fact that the definition of barred galaxies could be different depending on the studies. The earlier visual inspections provided us with the bar frequency of about 60\% including SB and SAB galaxies \citep{1973ugcg.book.....N, 1987rsac.book.....S, RC3, 2015ApJS..217...32B, Ann15}, while most of the recent visual inspections suggest $\sim$30\% of bar fraction, which is consistent with the fraction of SB galaxies \citep{2010ApJ...714L.260N, 2011MNRAS.411.2026M, 2011MNRAS.415.3627H, 2012ApJS..198....4O, 2012ApJ...745..125L, 2012MNRAS.423.1485S, 2013ApJ...779..162C, 2014MNRAS.445.3466S}. Among the automatic methods, ellipse fitting method yields a level of $\sim$45\% for the bar fraction \citep{2007ApJ...659.1176M, 2010ASPC..432..219M, 2007AJ....133.2846R, 2008ApJ...675.1194B, Agu09}, while Fourier analysis could not show consistent result for the bar fraction among the studies which used different criteria \citep{1990ApJ...357...71O, 2000AA...361..841A, Lau02, 2004ApJ...607..103L, Agu09}. 

In this paper, we introduce our sample, data reduction, and bar classification methodology in Section \ref{chap2}. Even though some studies used the same basic method to classify barred galaxies, they adopted different criteria. In Section \ref{chap3}, we compare the bar classification result obtained by applying three methods and diverse criteria for the same sample. We report in Section \ref{chap4} how the overall bar fractions are different depending on the classification methods and selection criteria despite being applied to the same sample. Next, we show that the dependence of the bar fractions on the properties of galaxies also become different depending on the classification method in Section \ref{chap5}. Finally, we compare our observational results with previous observational and theoretical studies in the context of bar formation and evolution in Section \ref{chap6}. The conclusion is given in Section \ref{chap7}. 

\section{Data Reduction and Bar Classification} \label{chap2}

\subsection{Sample and Data Reduction}\label{chap2.1}
We used the catalog of \citet[hereafter Ann15]{Ann15} to compare automatic classifications of barred galaxies against visual classification. The catalog provides detailed morphological types based on visual inspections for 5,836 galaxies from the SDSS/DR7 with $z < 0.01$. It contains 1,876 spiral galaxies ($\sim$32.1\%) classified according to the classification system of the Third Reference Catalog of Bright Galaxies \citep[hereafter RC3]{RC3}. Spiral galaxies are classified into three classes, normal spirals (SA), weak bars (SAB), and strong bars (SB), and ten stages from S0/a to Sm for the Hubble type. This detailed classification enable us to consider not only strong bars but also weak bars which have been excluded in most of recent studies and investigate the bar fraction as a function of Hubble sequence.

We made a volume-limited sample of 1,698 objects brighter than $M_{\rm r} = -15.2$ mag. After excluding edge-on galaxies identified in \citetalias{Ann15}, we obtained $g-$ and $i-$band images for 1,163 objects from the SDSS/DR7. All images went through three processes to subtract the soft-bias, sky and bright regions such as adjacent galaxies, foreground/background stars, and stellar clumpy regions within the target galaxies for automatic analysis. Eliminating sky background gradient is important in determining the position angle and ellipticity of the disk, needed for deprojection. 

\begin{figure*}[htbp]
\includegraphics[bb = 0 60 580 420, width = 0.95\linewidth, clip = ]{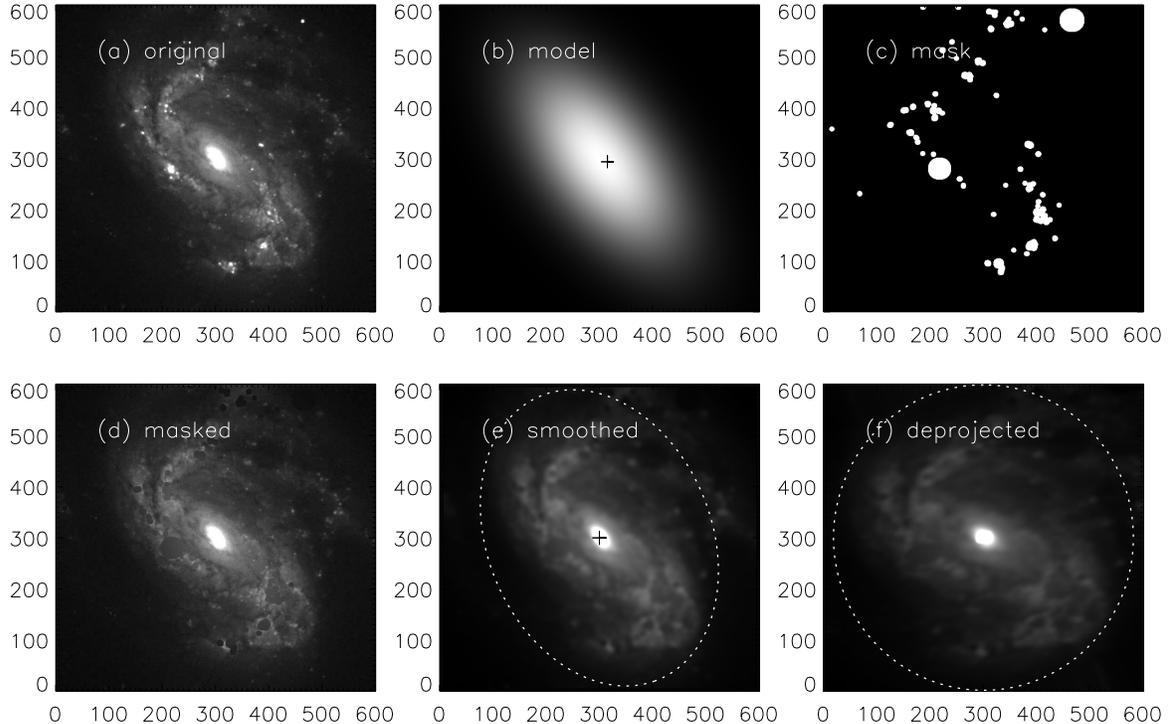}
\caption{Reduction and deprojection processes. (a): original image, (b): gaussian model of the target image overplotted the center, (c): masking filter, (d): subtracted image where the bright clumps are replaced with the model values, (e): smoothed image, (f): deprojected image. We overlaid the ellipse on $R_{25}$ in (e) and (f). \label{fig2.2}}
\end{figure*}

In particular, our sample galaxies are the nearest ones with $z < 0.01$, so they show a lot of resolved HII regions as well as foreground stars (Fig. \ref{fig2.2}a). We have to mask these bright clumpy sources to facilitate the automatic classification procedures. We started with analyzing the light distribution of the original image and determined a gaussian model for the target galaxy (Fig. \ref{fig2.2}b). Next, bright clumpy sources have been found (Fig. \ref{fig2.2}c) and replaced with the value given by the fitted model (Fig. \ref{fig2.2}d). We applied automatic reduction procedure to all target galaxies, and after visual check, we amended the reduction using interactive parameters for $\sim$10\% of objects for which automatic reductions were not sufficient. In addition, we smoothened images (Fig. \ref{fig2.2}e) after masking process. We chose smoothing boxes with a size of 0.1$R_{\rm 25}$, which best suited to our analysis. This smoothing process has not been applied in previous studies, which used smaller higher-redshift images or lower resolution images. We will discuss the effect of smoothing process on the results in \S \ref{chap3.2}. 

We deprojected galaxies to the face-on orientation assuming that the disk has a perfect circular shape. The position angle and ellipticity have been determined at the isophote of 25 mag/arcsec$^2$ in $B-$band \citep{2002ApJ...567...97L, 2002ApJ...575..156J}. We constructed $B-$band images by the relation $m_{\rm B} = m_{\rm g}+0.3$ \citep{2006PhRvL..97i7801R} from SDSS $g-$band images. We confirmed that the relation $m_{\rm B} = m_{\rm g} + 0.3130\times(m_{\rm g}-m_{\rm r}) + 0.2271$ \citep{2005AAS...20713308L} also produces nearly the same results. We evaluated the position angle and ellipticity by fitting the ellipse in IDL as mentioned in \S \ref{chap2.2.1}. We present the final deprojected image in Fig. \ref{fig2.2}f. After deprojection, we excluded from our sample 195 galaxies with inclination higher than 60 degrees and 84 galaxies with frames smaller than $R_{\rm 25}$. Ultimately, we have a subsample of 884 galaxies in the magnitude range $-21 \le M_r \le -15$ suitable for the automatic analysis.

\subsection{Bar Classification Methodology}\label{chap2.2}

\subsubsection{Ellipse Fitting}\label{chap2.2.1}
\citet{1995AAS..111..115W} showed that the characteristics of bar signature are increasing ellipticity ($\epsilon$) and constant position angle (PA) in the radial profiles derived from ellipse fits. \citet{2004ApJ...615L.105J} proposed two criteria to identify bars: (1) $\epsilon$ must rise to a global maximum above 0.25 ($\epsilon_{\rm bar}\geq0.25$) while the PA remains relatively constant within $\pm$20 degrees along the bar ($\Delta \rm PA_{bar}\leq \pm 20^\circ$), (2) the transition occurs between the bar and the disk region with $\epsilon$ dropping by more than 0.1 ($\Delta\epsilon_{\rm tra} \geq0.1$) and with PA changing by more than 10 degrees ($\Delta\rm PA_{tra} \geq 10^\circ$). Therefore, this bar selection method is defined by criteria on four parameters, the ellipticity threshold for bar ($\epsilon_{\rm bar}$), the range of constant PA over bar region ($\Delta \rm PA_{\rm bar}$), and the change of ellipticity ($\Delta\epsilon_{\rm tra}$) and PA ($\Delta \rm PA_{\rm tra}$) at the transition between a bar and a disk. The threshold for bar ellipticity distinguishes bars from oval structures and the constant PA distinguishes bars from spiral arms.

These criteria have been widely used with some variations. We list the criteria and the bar fraction results from previous studies in Table \ref{Table1}. We want to emphasize that even if we use the ellipse fitting method to detect bars, the results may depend on the different criteria adopted. In fact, we see wide range of the bar fractions from previous studies in Table \ref{Table1}, even after considering the fact that they are affected by the different wavelength or sample properties.

 Besides, \citet{2012ApJ...746..136M} tested additional criteria for bars whether $\epsilon_{\rm bar}$ should be a global maximum or not. \citet{2007ApJ...659.1176M} compared the difference between observed images and deprojected images. We tested various criteria and conditions such as wavelength and deprojection effect in order to find the optimal conditions to identify bars. 

We calculated the ellipse fits following \citet{1985AJ.....90..169D} and \citet{1990MNRAS.245..130A} in IDL. It is also based on the Fourier decomposition along a given ellipse, $(x^2/a^2)+(y^2/b^2) = 1$, where x is in the direction of the major axis: 
 
\begin{eqnarray}\label{eq1}
I(a,\phi) = I_{0}(a) + \sum_{m=1}^{\infty}[A_{m}(a) \cos m\phi + B_{m}(a) \sin m\phi],
\end{eqnarray}
where
\begin{eqnarray}
A_{m}(a) = 1/\pi \int_{0}^{2\pi}I(a,\phi)\cos m\phi d\phi,
\end{eqnarray}
and
\begin{eqnarray}
B_{m}(a) = 1/\pi \int_{0}^{2\pi}I(a,\phi)\sin m\phi d\phi.
\end{eqnarray}

We analyzed the luminosity profiles $I(a,\phi)$ constructed from initial parameters such as ellipticity ($\epsilon$), position angle ($\theta$) and center position ($x_c$ and $y_c$). From the initial solution, we obtained the Fourier coefficient $A_1, B_1, A_2$, and $B_2$, which are related to $x_c$, $y_c$, $\epsilon$, and PA \citep{1979ApJ...234...76Y, 1983ApJ...266..562K}. We adjusted the ellipse parameters that yield the largest $A_m$ or $B_m$ and accepted the final isophote when $|A_m| < 10^{-3}$ and $|B_m| < 10^{-3}$. We confirmed that our results are consistent with those from IRAF/ellipse package based on \citet{1983ApJ...266..562K} and \citet{1987MNRAS.226..747J} who directly correct the ellipse parameters at each iteration by using the relation between coefficients and parameters. Our method converges to robust results, while being less affected by initial parameters compared to the IRAF/ellipse package. 

We performed ellipse fitting with step sizes increasing by 10\% from three pixels from the center. The center was taken as the centroid within a given isophote, generally the isophote at the intensity 1/3 of the peak intensity. We used the fixed center for all radii. Fitting with a fixed center is more efficient in finding the transition between a bar and a disk, although varying center may reflect the real light distribution better. Our code finds transitions between bars and disks at first, and then checks if the bar candidates satisfy the criteria for ellipticity and PA.

\startlongtable
\begin{deluxetable*}{l|lccccll} 
\tablecaption{Different criteria for bar detection in ellipse fitting method and the resulting bar fraction $F_{\rm bar}$ \label{Table1}}
\tablehead{ 
  \colhead{     Author   }             &
  \colhead{$\Delta\epsilon_{\rm tra}$}         &
  \colhead{$\Delta \rm PA_{\rm tra}$}               &
  \colhead{$\Delta \rm PA_{bar}$}         &
  \colhead{$\epsilon_{\rm bar}$} &
  \colhead{$\rm F_{\rm bar}$(\%)}        &
  \colhead{wavelength}          \\ 
  \colhead{(1)} &
  \colhead{(2)} &
  \colhead{(3)} &
  \colhead{(4)} &
  \colhead{(5)} &
  \colhead{(6)} &
  \colhead{(7)} \\
}
\startdata
\citet{1995AAS..111..115W} & 0.02 & 2$^\circ$  & const & - & -  & $B, V, R, I$ \\
\citet{2002ApJ...567...97L} & 0.1  & -          & $\pm10^\circ$ & - &   17 & NIR  \\
\citet{2004ApJ...615L.105J} & 0.1  & $10^\circ$ & $\pm20^\circ$
                            & $\epsilon_{max} > 0.4$ &   33 & optics\\
                            &      &            &               &                      &   36 & NIR  \\
\citet{2007ApJ...657..790M} & 0.1 & $10^\circ$ & const & $\epsilon_{max}> 0.2$
                            &   59 & $J+H+K$  \\
\citet{2007ApJ...659.1176M} & 0.1 & change & $\pm5^\circ$ & $> 0.25$
                            &   $44\pm7$ & optics\\
                            &     &        &               &
                            &   $60\pm7$ & NIR \\
\citet{2007AJ....133.2846R} & & & & &   47 & $I-$band \\
\citet{2008ApJ...675.1194B} & 0.1 & change & $\pm5^\circ$
                            & $\epsilon_{max} >0.25$ &   48 & $r-$band \\
\citet{2008ApJ...675.1141S} & 0.1 & 10$^\circ$ & const  & $ > 0.2$ &   65 & NIR \\
\citet{Agu09}               & 0.08 & 5$^\circ$ & $\pm10^\circ$ & - &    45 & $r-$band \\
\citet{2009ApJ...698.1639M} & 0.1 & change & $\pm10^\circ$
                            & $\epsilon_{max}>0.25$ & $  \sim30$
                            & optics\\
\citet{2010ASPC..432..219M} & 0.1 & change & $\pm10^\circ$ &
                            $\epsilon_{max}>0.25$ &   $47\pm11$ & NIR \\ 
\citet{2012ApJ...746..136M} & 0.1 & 10$^\circ$ & $\pm10^\circ$
                            & $\epsilon_{max}>0.25$ &   $50\pm11$ & NIR \\
                            & & & & $ > 0.25$ &   $65\pm 11$ & \\ 
\citet{2016AA...595A..67C} & 0.08 &  20$^\circ$    & $\pm20^\circ$  
                           & -   &   36 & optics \\
\enddata
\tablenotetext{1}{Notes- Col. (1): Previous studies using ellipse fitting method. Col. (2): $\epsilon$ transition criterion between bars and disks. The values of \citet{1995AAS..111..115W} in Col. (2) and (3) are mean differences at transition, not criterion. Col. (3): PA transition criterion between bars and disks. Col. (4): Range of constant PA over bar region to distinguish between bars and spiral arms. Col. (5): Limit on $\epsilon$ to distinguish bars from oval structures. In particular, $\epsilon_{\rm max}$ indicates that the ellipticity of bar should be the global maximum. Jogee et al. (2004) suggested the criterion for weakly bars as $0.25 < \epsilon_{\rm max} \leq 0.4$. Col. (6): Resulting bar fraction from each research. Col. (7): wavelength.}

\end{deluxetable*}

\subsubsection{Fourier Analysis}\label{chap2.2.2}

\begin{deluxetable*}{l|lcccc}
\tablecaption{Bar fractions, different criteria for the Fourier analysis \label{Table2}}
\tablehead{
  \colhead{Author}  & 
  \colhead{$I_{m}/I_{0}$} & 
  \colhead{$\Phi_{m}$} &
  \colhead{$F_{bar}(\%)$} &
  \colhead{$F_{bar}$($z < 0.01$)(\%)}\\
  \colhead{(1)} &
  \colhead{(2)} &
  \colhead{(3)} &
  \colhead{(4)} &
  \colhead{(5)} \\
}
\startdata
\citet{1990ApJ...357...71O} & $I_{b}/I_{ib} > 2$ & - & - & 72.29 \\
\citet{2000AA...361..841A} & $I_{b}/I_{ib} > (max-min)/2+min$ &  -  & - & 51.47 \\
\citet{Lau02} & $I_{2}/I_{0} > 0.3$ & const $\Phi_{2}$, $\Phi_{4}$ & 40 & 36.31 \\
\citet{2004ApJ...607..103L} & $A_{2} > 0.12$ (SB) & const $\Phi_{2}$, $\Phi_{4}$ & 65 & 32.47 \\
               & $A_{2} >0.09$  (SAB) &      &      &  &   \\          
\citet{Agu09}  & $\Delta(I_{2}/I_{0}) \geq 0.2$ &  constant $\Phi_2$  & 26 & 53.85 \\
               &  &  $(\Delta\Phi_{2} < 20^\circ)$ &  &  \\
\enddata
\tablenotetext{1}{Notes- Col. (1): Previous studies using Fourier analysis. Col. (2): Criteria for relative Fourier amplitude. Col. (3): Criteria for constant phase angle. Col. (4): Bar fraction from previous studies. Col. (5): Bar fraction obtained in our study by applying each criterion.} 
\end{deluxetable*}

An alternative way to classify bars automatically is to utilize the Fourier coefficient directly \citep{1990ApJ...357...71O, 1998AJ....116.2136A, 2000AA...361..841A}. Previous studies have used the deprojected images and analyzed the azimuthal luminosity profiles $I(r,\theta)$ along concentric ellipses with Fourier decomposition as shown in equation (\ref{eq1}).

To identify bars, some studies \citep{1990ApJ...357...71O, 2000AA...361..841A} have used the ratio of bar intensity to interbar intensity, $I_{\rm b}/I_{\rm ib}$, where the bar intensity $I_{\rm b}$ is defined as the sum of the even Fourier components, $I_{\rm 0}+I_{\rm 2}+I_{\rm 4}+I_{\rm 6}$, and the inter-bar intensity $I_{\rm ib}$ is given by $I_{\rm 0}-I_{\rm 2}+I_{\rm 4}-I_{\rm 6}$. On the other hand, others \citep{Lau02, 2004ApJ...607..103L, Agu09} used the relative Fourier amplitude of $m$-th component described as

\begin{eqnarray}
\frac{I_{\rm m}(r)}{I_{\rm 0}(r)} = \frac{[A_{\rm m}(r)^2+B_{\rm m}(r)^2]^{1/2}}{A_{\rm 0}(r)/2}.
\end{eqnarray}
Generally, even terms, in particular m = 2, appear much larger than odd terms over the bar region \citep{1990ApJ...357...71O, 2000AA...361..841A}. However, other non-axisymmetric structures such as spiral arms can produce the same effect. Therefore, some studies also demanded the phases defined by  
\begin{eqnarray}
\Phi_{\rm m}(r) = \tan^{-1} \frac{B_{\rm m}(r)}{A_{\rm m}(r)}
\end{eqnarray}
to remain constant in order to distinguish bars from spiral arms \citep{Lau02, 2004ApJ...607..103L, Agu09}. In particular, \citet{Agu09} utilized these parameters to investigate the transition between a bar and a disk as the way in the ellipse fitting method. We list their criteria for bar classification and the bar fractions they obtained in Table \ref{Table2}.

We reproduced all of their bar classifications using deprojected $i-$band images as in previous studies and compared each other. However, we did not exactly reproduce \citet{2004ApJ...607..103L}. They performed two-dimensional bar-bulge-disk decomposition before deprojection process and applied the deprojection only to disk and bar component because spherical bulge is not, in principle, affected by projection. However, paradoxically, it needs the information about bar structure to decompose the image into bar, bulge and disk before bar classification. They considered both cases in which galaxies host a bar or not and we just applied their criterion to wholly deprojected image. 

\section{COMPARISON OF BAR CLASSIFICATION}\label{chap3}

\subsection{Visual Inspection}\label{chap3.1}
We compared the results of the bar classifications by different methods but applied to the same sample. At first, we compared the classifications by two independent visual inspections, \citetalias{Ann15} and \citetalias{RC3} catalog. \citetalias{Ann15} classified 5,836 galaxies with Petrosian $\it r-$magnitudes brighter than 17.77 with $z < 0.01$. \citetalias{RC3} catalog classified more than 23,000 galaxies larger than one arc minute and brighter than $B = 15.5$ with $z < 0.05$. \citetalias{Ann15} classification system follows that of \citetalias{RC3} classification. Both have classified spiral galaxies as not only SA and SB but also SAB. \citetalias{Ann15} found 361 strong bars (SB, 31.0\%) and 365 weak bars (SAB, 31.4\%) out of final 1,163 disk galaxies after rejecting 535 edge-on galaxies. \citetalias{RC3} catalog provides the classification for 1,707 spirals within $z = 0.01$. It comprises SA of 24\%, SAB of 28\% and SB of 48\% for 1,274 galaxies after excluding 433 uncertain or doubtful objects to determine whether they host a bar or not. We found that two catalogs include 706 common spiral galaxies.

\begin{figure}[htbp]
\includegraphics[bb = 120 380 440 650, width = 0.9\linewidth, clip = ]{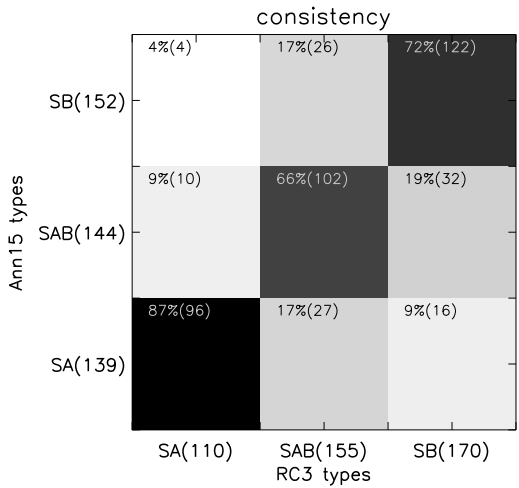} 
\caption{Comparison of the morphological types of RC3 vs. Ann15. The numbers indicate matched percentages of classification of Ann15 (y-axis) for classification of RC3 (x-axis). It shows the consistency between two catalogs after rejecting unclassified objects such as edge-on and uncertain galaxies. The darker shade means higher agreement. \label{fig3.1}}
\end{figure}

We investigated the agreement of classification for these 706 galaxies between two catalogs. In Fig. \ref{fig3.1}, we present the percentage of classification of \citetalias{Ann15} (y-axis) against the classification of \citetalias{RC3} (x-axis). The higher number and darker shade mean better agreement. We confirmed that the mutual concordances reach up to 87\%, 66\%, and 72\% in SA, SAB, and SB categories, respectively, after excluding edge-on and uncertain galaxies in each catalog. \citetalias{Ann15} classified only 4\% (4 galaxies out of 111) of SAs determined by \citetalias{RC3} as SBs and \citetalias{RC3} categorized just 9\% (16 galaxies) of SAs by \citetalias{Ann15} as SBs. Therefore, the disagreement between SA and SB galaxies is less than 10\%. However, when it comes to weak bars, we found that the agreement is not as good. Out of 155 SABs from \citetalias{RC3}, 102 galaxies (66\%) have been classified as SAB by \citetalias{Ann15}. Remaining 53 SABs in \citetalias{RC3} are evenly classified either as SA (27 galaxies) or SB (26 galaxies) in \citetalias{Ann15}. This is in a way expected because SAB is intermediate between SA and SB, and its classification would include more ambiguity compared to SA or SB.

We will use only the matched galaxies between two studies to compare with automatic classification in \S\ref{chap3.2} and \S\ref{chap3.3}. Rejecting highly inclined galaxies and objects with a smaller image size than $R_{\rm 25}$ leaves only 211 galaxies. Therefore, the galaxies which we use to compare with automatic methods are 63 SA, 64 SAB, and 84 SB galaxies, which we call `concordance sample'.

\subsection{Ellipse Fitting Method}\label{chap3.2}

We compare the classification by ellipse fits versus that by visual inspection using the concordance sample discussed above. As mentioned before, ellipse fitting methods have been applied with different criteria depending on studies. In this paper, we tested each criterion in order to find the optimal criteria that yield the highest agreement with visual classification. We experimented with transition threshold ($\Delta \rm PA_{\rm tra}$), ellipticity limit ($\epsilon_{\max}$), deprojection, wavelength band, and smoothing box size. 

\begin{figure*}[htbp]
\includegraphics[bb = 0 410 480 610, width = 0.93\linewidth, clip=]{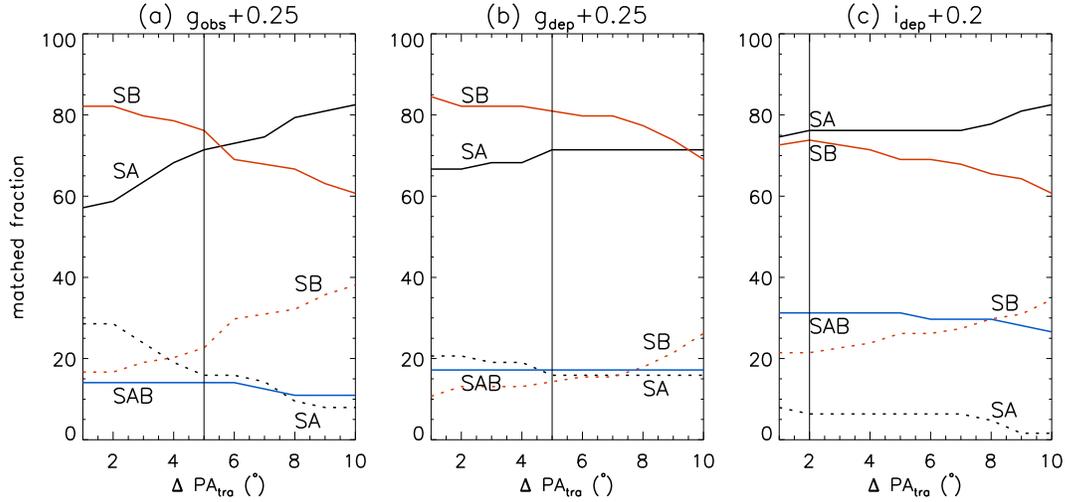}
\caption{matched (solid lines) and unmatched (dotted lines) fractions between classifications by ellipse fits and visual inspection. They become different depending on the criteria. SB and SAB are defined as $\epsilon_{\rm bar}\geq$ 0.4 and 0.25 (or 0.2) $\leq \epsilon_{\rm bar} <$ 0.4, respectively, following \citet{2004ApJ...615L.105J}. The $\Delta \epsilon_{\rm tra}$ with 0.1 and constant $\rm PA_{\rm bar}$ within $\pm$ 10$^\circ$ are applied to all cases. Each panel shows the matched and unmatched fractions as a function of $\Delta \rm PA_{\rm tra}$ in different conditions: (a) $g-$band observed images with $\epsilon_{\rm bar}\geq$ 0.25; (b) $g-$band deprojected images with $\epsilon_{\rm bar}\geq$ 0.25; (c) $i-$band deprojected images with $\epsilon_{\rm bar}\geq$ 0.2 \label{fig3.2.1}}
\end{figure*}

For the transition between the bar and the disk, most studies adopted 0.1 as $\Delta\epsilon_{\rm tra}$, but a different value for $\Delta\rm PA_{\rm tra}$. Whereas some studies used 10$^\circ$ or 5$^\circ$, others did not specifically constrain $\Delta \rm PA_{\rm tra}$ and consider any change in PA profile as a signature of a bar. In Fig. \ref{fig3.2.1}, we show how the value of $\Delta \rm PA_{\rm tra}$ affects the agreement and disagreement with visual classification. The matched percentage of classification by ellipse fitting versus visual classification are displayed as a function of $\Delta \rm PA_{tra}$. The solid lines and dotted lines show the agreement and disagreement with the visual classification, respectively. Although we distinguish weak bars from strong bars by their $\epsilon_{\rm bar}$, $0.25 \le \epsilon_{\rm bar} < 0.4$ for SAB and $\epsilon_{\rm bar} \geq 0.4$ for SB \citep{2004ApJ...615L.105J}, the agreement rate of SAB with visual classification is very low. 

\begin{figure*}[htbp]
\includegraphics[bb = 0 610 600 800, width = 0.85\linewidth, clip=]{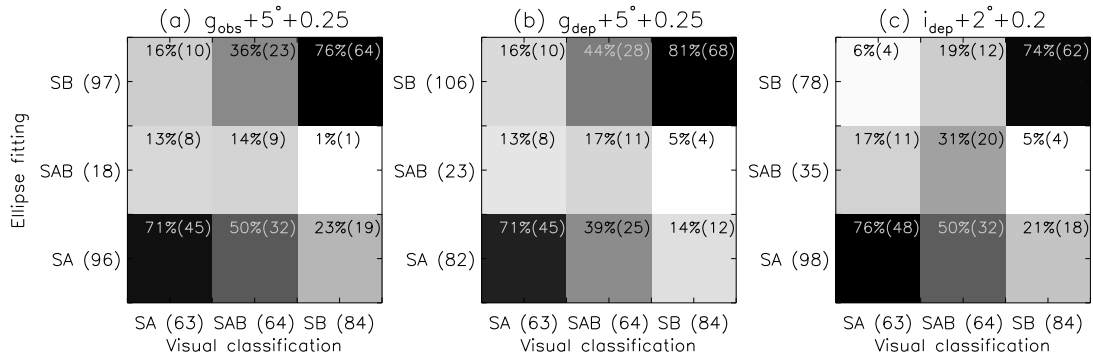}
\caption{Comparison between automatic classification by ellipse fitting and the visual classification. The numbers represent the matched percentages of ellipse fits (y-axis) as a function of visual inspection (x-axis). The optimal criterion of $\Delta \rm PA_{\rm tra}$ found from Fig. \ref{fig3.2.1} is applied to each condition: (a) $\Delta \rm PA_{\rm tra} \ge 5^\circ$ in $g-$band observed images; (b) $\Delta \rm PA_{\rm tra} \ge 5^\circ$ in $g-$band deprojected images; (c) $\Delta \rm PA_{\rm tra} \ge 2^\circ$ in $i-$band deprojected images \label{fig3.2.2}}
\end{figure*}

We found that as we apply higher threshold for $\Delta \rm PA_{\rm tra}$, the agreement of SA (black solid line) increases whereas those of SB (red solid line) and SAB (blue solid line) decrease. It is because a large $\Delta \rm PA_{\rm tra}$ would classify an aligned bar with a disk as a non-barred galaxy as noted earlier \citep{2007ApJ...657..790M}. However, if we deproject the images, we can reduce the effect of $\Delta \rm PA_{\rm tra}$ as shown in Fig. \ref{fig3.2.1}b and \ref{fig3.2.1}c. Using deprojected images in general increases the match between the ellipse fitting method and visual classification and also lessen the dependence on the choice of $\Delta \rm PA_{\rm tra}$. We found the optimal threshold of $\Delta \rm PA_{\rm tra}$ in each case to increase the matched fraction and reduce the unmatched fraction for SB and SA against visual classification. We indicate them as black vertical lines in Fig. \ref{fig3.2.1} and present their consistency against visual classification in Fig. \ref{fig3.2.2} in the same way as in Fig. \ref{fig3.1}.

In general, the ellipse fitting method classifies around 70\% of visually determined SAs as SAs and 70\% $\sim$ 80\% of visually classified SBs as SBs. The ellipse fitting method seems to be as effective in detecting barred galaxies as two independent visual inspections described in \S\ref{chap3.1}. However, the confusion between SB and SA in ellipse fitting method is somewhat higher than that in visual classification. We investigated missed SBs in the automatic classification and found that they are mostly early-type spirals such as S0/a, Sa, or Sab. A large bulge makes the ellipticity of bar lower and dilutes the transition between a bar and a disk in ellipticity profiles. \citet{Agu09} has noted earlier that the ellipse fitting method detects more efficiently bars with a sharp end than those with a smooth transition by testing artificial galaxies. On the other hand, it sometimes classifies tightly wound spiral arms as bar structures and they are usually Sc or Scd. Therefore, the ellipse fitting method can affect the bar fraction differently for different Hubble types as will be discussed in \S \ref{chap5.2}.

Deprojecting $g-$band images can increase the agreement for SBs by 5\% and reduce the missed SBs by 9\% (Fig. \ref{fig3.2.2}b). Deprojection process somewhat helps disentangle the aligned bar from the inclined disk. It also helps a little find bars in early-type spirals with a large bulge. Therefore, using $g-$band deprojected images with the criterion of $\Delta \rm PA_{\rm tra} \ge 5^\circ$ and $\epsilon_{\rm bar}\geq$ 0.25 seems to be the optimal choice to construct the ellipse fitting method most consistent with the visual classification. 

We had expected that $i-$band images are more favorable to detect bars because NIR images not only reflect old populations that make up the bar structure but also reduce the dust obscuration. Moreover, previous studies had presented higher bar fraction in NIR \citep{2000AJ....119..536E, 2007ApJ...659.1176M, 2015ApJS..217...32B}. However, we found that using $i-$band images decreases the matched fraction of SB while increasing that of SA (Fig. \ref{fig3.2.1}c). It is because $i-$band images have smoother light distribution than $g-$band images and it works unfavorable for the algorithm to find the transition between a bar and a disk. We obtained similar agreement with using $g-$band observed images when we applied lower threshold of $\Delta \rm PA_{\rm tra} \ge 2^\circ$ and $\epsilon_{\rm bar} \ge 0.2$ in $i-$band deprojected images (Fig. \ref{fig3.2.2}c). 

We experimented with the lower limit of 0.25 and 0.2 for the $\epsilon_{\rm bar}$ in all cases. Lower threshold helps detect more visually determined SABs, but, at the same time, has a risk of misjudging bulges of nonbarred galaxies as weak bars. For the visual classification, our eyes distinguish bulges from weak bars considering not only the ellipticity but also other characteristics such as the light distribution of bulge. Only for $i-$band images, the lower limit of 0.2 helps increase both of the matched fractions for SA and SAB. That's why we selected different criteria of $\epsilon_{\rm bar}$ for $g-$ and $i-$band images. 

In addition, we tested the effect of smoothing box size. Although we need to smoothen images in order to minimize the effect of the artificial residuals after masking, we caution that larger smoothing box can hide the transition between a bar and a disk. It is similar to that lower angular resolution by the limited FWHM of the telescope or from the distance to the galaxies would decrease the detection of bars. In our test, smaller smoothing box increases the agreement of SBs with visual inspection, while decreases that of SAs. We found that the size of $0.1R_{25}$ is optimal for our sample galaxies in order to increase both of the matched fraction for SB and SA.

\begin{deluxetable*}{l|rrr|rrr|rrr}[tbp]
\tablecaption{Comparison between the classifications by visual inspection and Fourier analysis \label{Table3.3}}
\tablehead{
  \colhead{visual class} & 
  \colhead{SA} &
  \colhead{ } & 
  \colhead{ } & 
  \colhead{SAB} &
  \colhead{ } & 
  \colhead{ } & 
  \colhead{SB} & 
  \colhead{ } & 
  \colhead{ } \\ 
  \colhead{Fourier class} & 
  \colhead{SA} &
  \colhead{SAB} &
  \colhead{SB} & 
  \colhead{SA} &
  \colhead{SAB} &
  \colhead{SB} & 
  \colhead{SA} &
  \colhead{SAB} &
  \colhead{SB} \\
}
\startdata
\citet{1990ApJ...357...71O} & 46.03 & 0. & 53.97 & 26.56 & 0. & 73.44 & 0. & 0. & 100. \\
\citet{2000AA...361..841A} & 57.14 & 0. & 42.86 & 50.00 & 0. & 50.00 & 34.52 & 0. & 65.48 \\
\citetalias{Lau02} & 79.37 & 0. & 20.63 & 56.25 & 0. & 43.75 & 15.48 & 0. & 84.52 \\
\citet{2004ApJ...607..103L} & 79.37 & 4.76 & 15.87 & 65.62 & 1.56 & 32.81 & 27.38 & 3.57 & 69.05 \\
\citet{Agu09}     & 46.03 & 0 & 53.97 & 31.25 & 0. & 68.75 & 11.90 & 0. & 88.10 \\
\enddata
\tablenotetext{1}{The numbers indicate the matched percentage of classification by Fourier analysis for that by the visual inspection. The weak bars have never been defined except for \citet{2004ApJ...607..103L}.}
\end{deluxetable*}

\subsection{Fourier Analysis}\label{chap3.3}

We also investigated the agreement between the automatic classification based on the Fourier analysis and that by visual inspection for the concordance sample. We applied five criteria listed in Table \ref{Table2} and summarized the matched percentage between the Fourier analysis and visual classifications in Table \ref{Table3.3}. 

The earlier criteria \citep{1990ApJ...357...71O, 2000AA...361..841A} constructed from very small samples seem to be very awkward to classify barred galaxies. They consider most of galaxies as barred galaxies or divide galaxies into half regardless of their morphologies. Their criteria are matched not only by bar structures but also by other diverse components. On the other hand, more recent studies which demand $\Phi_2$ and $\Phi_4$ being constant over a bar region greatly improve the agreement with the visual classification. In particular, the criterion of \citet[hereafter Lau02]{Lau02} yields the outstanding result in distinguishing barred galaxies from non-barred galaxies. It measures non-barred galaxies and barred galaxies with a matched fraction of 79\% and 87\% with visual inspections, respectively. It is the highest agreement with visual inspections among all of the methods and criteria we have tested. Nevertheless, they missed eleven visual SB galaxies (13\%). All of them have gradually increasing or decreasing phase profiles ($\Phi_{m}(r)$) caused by a large bulge or spirals wrapped with the bar. On the other hand, they classified 13 visually selected SAs (21\%) as barred galaxies. We found that three of them have bar signatures which previous visual inspections did not find. The rest of misjudged SBs are located near the thresholds of criteria, and half of them are early-type spirals with a large bulge. The inclined bulges are often confused with a bar. In other words, misclassifications by Fourier analysis against visual classifications are increasing in early-type spirals. A large bulge is very tricky to deal with in automatic classification and our deprojection process can not solve the inclination problem completely. 

\citet{2004ApJ...607..103L} used $A_2$ instead of relative amplitude, and it missed more visually classified strong bars (29\%) than before (13\%). It is because many visual SBs, especially late-type spirals such as Sd, do not have $A_2$ larger than the threshold suggested by \cite{2004ApJ...607..103L}, despite satisfying the criterion on $I_2/I_0$ used by \citetalias{Lau02}. Therefore, the relative amplitude would be more useful in determining a bar structure than the absolute amplitude. On the contrary, the criterion of \citet{Agu09} distinguishes only 46\% of visually non-barred galaxies from barred galaxies, whereas it shows the high agreement of 87\% with visual classifications for SBs. We found that it is because they didn't set any threshold on $I_2/I_0$ for bar structures to be distinct from the inclined bulge. In practice, criterion by \citet{Agu09} classifies many inclined bulges or weakly oval structures which have a constant phase as bars. 

\section{BAR FRACTION}\label{chap4}

\begin{deluxetable}{l|ccc}[b]
\tablecaption{Bar fraction by different classification methodology \label{Table4}}
\tablehead{
 \colhead{classification method} &
 \colhead{$F_{\rm bar}$} &
 \colhead{$F_{\rm SB}$} &
 \colhead{$F_{\rm SAB}$} \\
}
\startdata
visual inspection \citepalias{Ann15}       & 63\% & 30\% & 33\% \\
ellipse fitting ($g_{\rm obs}$+5$^\circ$+0.25) & 48\% & 41\% & 7\% \\
ellipse fitting ($g_{\rm dep}$+5$^\circ$+0.25) & 56\% & 45\% & 11\% \\
ellipse fitting ($i_{\rm dep}$+2$^\circ$+0.20) & 52\% & 35\% & 17\% \\
Fourier analysis \citepalias{Lau02} & 36\% & 36\% & - \\
\enddata
\tablenotetext{1}{Most bars calculated by ellipse fits fall under the strong bars, $\epsilon_{\rm bar} \geq 0.4$, and a small fraction stays under $0.25 \leq \epsilon_{\rm bar} < 0.4$ for weak bars.}
\end{deluxetable}

We measured the overall bar fractions from the final 884 sample galaxies using visual inspections, ellipse fits, and Fourier analysis. The results are listed in Table \ref{Table4}. The resulting bar frequencies become different depending on the methods or criteria to detect bars even for the same sample galaxies. Furthermore, wavelength band of images, deprojection, and spatial resolution also influence the fraction of detected bars. It is apparent why researchers derived different frequencies of barred galaxies.

We calculated the bar fraction which includes strong bars (30\%) and weak bars (33\%) from visually determined catalog of \citetalias{Ann15}. It is similar to the typical bar fraction of classical visual classification such as UGC, RSA \citep{1973ugcg.book.....N, 1987rsac.book.....S} and \citetalias{RC3}, which present the fraction of $\sim$30\% for SB and SAB, respectively. \citet{2015ApJS..217...32B} also obtained similar bar fraction of 66\% including SAB by visual inspection of 1160 galaxies in NIR wavelength. On the other hand, many of recent visual inspections show a frequency around 30\% \citep{2010ApJ...714L.260N, 2011MNRAS.411.2026M, 2011MNRAS.415.3627H, 2012ApJS..198....4O, 2012ApJ...745..125L, 2012MNRAS.423.1485S, 2013ApJ...779..162C, 2014MNRAS.445.3466S}. They usually deal with obvious bars only, and almost all of weak bars may have been excluded. Their bar fractions are consistent with the fraction of strong bars in classical visual classification.

From the ellipse fitting method, we obtained the bar fraction in the range of 48\%$\sim$56\%, depending on the detailed criteria or conditions. It is similar to the ranges of previous studies by ellipse fits \citep{2007ApJ...657..790M, 2007ApJ...659.1176M, 2010ASPC..432..219M, 2012ApJ...746..136M, 2007AJ....133.2846R, 2008ApJ...675.1194B, Agu09}. The fraction of detected bars is higher than the fraction of strong bars only, yet less than that of all bars, strong and weak, from visual inspection. In other words, the fraction of bars can be different depending on how many weak bars are included. The most typical condition in previous ellipse fitting method, $g_{\rm obs}+5^\circ+0.25$, yields 48\%, the most typical bar fraction of previous studies by ellipse fits, which is about 50\% larger than the fraction of the strong bars only by visual inspection. In fact, this condition $g_{\rm obs}+5^\circ+0.25$ classifies about half of visual SAB as barred galaxies and rest of them as non-barred galaxies (Fig. \ref{fig3.2.2}a). And we obtained the closest bar fraction to visual inspection when we use the condition $g_{\rm dep}+5^\circ+0.25$. The bar fraction is 56\% and it classifies about 60\% of visual SABs as SAB or SB (Fig. \ref{fig3.2.2}b).
  
Table \ref{Table4} shows a lower bar fraction from the $i-$band images than from the $g-$band images, contrary to previous studies which obtained higher bar fraction in NIR (60\%) than in optical (44\%) by ellipse fitting method from HST \citep{2007ApJ...659.1176M}. When we adopted the condition of $i_{\rm dep}+3^\circ+0.2$ with smaller smoothing box of $0.05R_{25}$, we also obtained a higher bar fraction of 56\%. The bar fraction does depend on the wavelength band or the spatial resolution of the galaxy images as discussed in previous studies \citep{2018Erwin}. 

Lastly, we obtained the bar fraction of 36\% from Fourier analysis using the criterion of \citetalias{Lau02} which shows the highest agreement with visual inspections. We confirmed \citetalias{Lau02} that also showed similar bar fraction of 40\% for sample within $cz < 2500 km s^{-1}$, which is similar to ours. The bar fraction from Fourier analysis is close to the fraction of SB obtained from visual inspection, and lower than those from the ellipse fitting method. 

Our analysis showed that we obtain different bar fractions depending on the method we applied. The difference is, firstly, due the ambiguity inherent in distinguishing weakly barred galaxies from non-barred galaxies, and, secondly, due the difficulty in identifying bars by automatic methods from large bulges or wrapped spiral arms as discussed in \S \ref{chap3}.

\section{BAR FRACTION AND HOST GALAXIES}\label{chap5}

Now, we investigate the dependence of bar frequency on the physical properties of host galaxies for each classification method. We use four parameters to represent the properties of galaxies: numerical code T, ($\it g-r$) color, fracdeV, and inverse light concentration $C_{\rm in}$. The parameter T is the indicator of the Hubble stage, its number running from 0 to 9 for S0/a, Sa, Sab, Sb, Sbc, Sc, Scd, Sd, Sdm, and Sm \citepalias{RC3}. They are grouped into early-type (S0/a-Sb), intermediate-type (Sbc-Scd), and late-type (Sd-Sm) spirals \citepalias{Ann15}. The SDSS parameter fracdeV is the fraction of the light fit by a de Vaucouleurs profile versus an exponential profile \citep{2010MNRAS.404..792M, 2011MNRAS.411.2026M}. Therefore, fracdeV indicates bulge-to-total ratio and large values mean bulge-dominated systems which have mostly classical bulges. Galaxies with small fracdeV are disk-dominated systems in which bulges are mostly pseudo-bulges or even bulgeless. The last parameter, inverse light concentration, is defined as $C_{\rm in}\equiv R_{\rm 50}/R_{\rm 90}$, where $R_{\rm 50}$ and $R_{\rm90}$ are the Petrosian radii enclosing 50\% and 90\% of the total galaxy light \citep{1976ApJ...209L...1P, 2012ApJ...745..125L}. It increases toward a less-concentrated system such as late-type spirals.

\begin{figure*}[htbp]
\includegraphics[bb = 0 420 590 820, width = 0.95\linewidth, clip=]{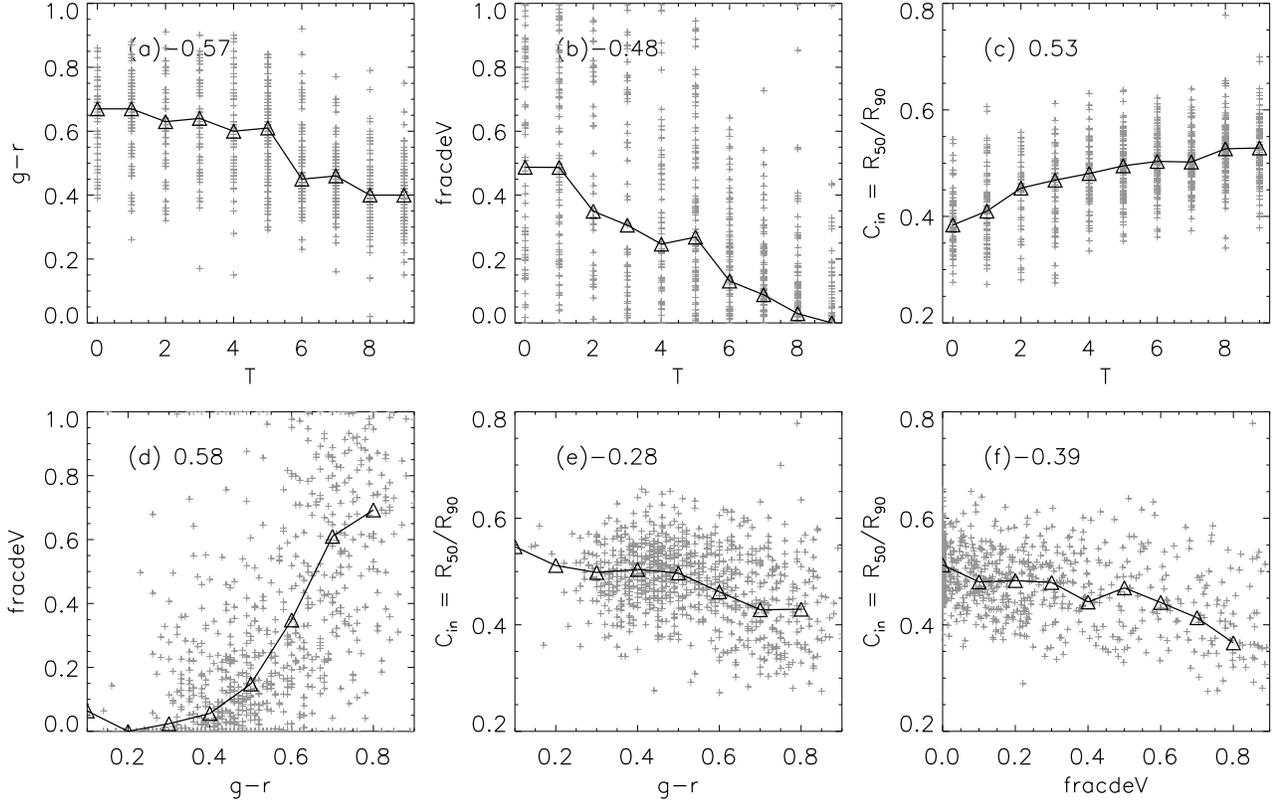} 
\caption{Correlation between properties of host galaxies. (a): T and $(\it g-r)$, (b): T and fracdeV, (c): T and $C_{\rm in}$, (d): $(\it g-r)$ and fracdeV, (e): $(\it g-r)$ and $C_{\rm in}$, (f): fracdeV and $C_{\rm in}$. Correlation coefficients are presented at the top left and triangles show the median values. \label{fig5}}
\end{figure*}

In fact, all these parameters are known to be related to the galaxy morphology \citep{1993IAUS..153..209K, 1996MNRAS.279L..47A, 2000ApJ...529..886C, 2001AJ....122.1707G, 2003ApJS..147....1C, 2005ApJ...635L..29P, Agu09}. We present the correlations among the parameters for our sample galaxies in Fig. \ref{fig5} and the correlation coefficients at the top left of each panel. Hubble type (T) shows comparatively good correlations with color, fracdeV, and $C_{\rm in}$ (Fig. \ref{fig5}a-c). Despite the large scatter, as T increases, the median values of ($\it g-r$) and fracdeV decrease, and $C_{\rm in}$ increases gradually. In particular, we note that early-type spirals (small T) span the whole range of fracdeV while late-type spirals a limited range of fracdeV (Fig. \ref{fig5}b). It implies that early-type spirals have not only classical bulges but also pseudo-bulges, whereas it is difficult for late-type spirals to have classical bulges. This is consistent with observational results that pseudo-bulges are discovered even in lenticular galaxies \citep{2009ApJ...692L..34L}. When we inspect the correlations among ($\it g-r$), fracdeV, and $C_{\rm in}$, we find that fracdeV and $C_{\rm in}$ are weakly anti-correlated with a coefficient of -0.39 even though both of them have often been used as an indicator of the bulge size (Fig. \ref{fig5}f). Also fracdeV has a relatively strong correlation with ($\it g-r$) which reflects stellar population or recent star formation \citep{2005ApJ...635L..29P}, while light concentration $C_{\rm in}$ has little correlation with $(\it g-r)$ (Fig. \ref{fig5}d-e).

Stellar mass is another important parameter that affects the bar formation and evolution \citep{2008ApJ...675.1194B, 2010ApJ...714L.260N, 2012ApJ...761L...6M, 2016A&A...587A.160D, 2018Erwin}. Our sample galaxies are distributed in the stellar mass range of $10^8 \le M_\ast/M_\odot \le 10^{11}$ following \citet{2003Bell}:
\begin{eqnarray}
\rm log\frac{\it M_{\ast}}{\it M_{\odot}} = -0.306+1.097(\it g-r \rm)-0.4(\it M_r-M_{r,\odot}).
\end{eqnarray}
However, we have to caution that this mass estimate may contain significant errors caused by the relatively large peculiar velocities, because our sample galaxies are within $z = 0.01$. 

 We used the value of T and $(\it g-r)$ from \citetalias{Ann15} catalog, and other information were obtained from the SDSS database. The color is corrected for the Galactic extinction.

\subsection{Bar Fraction by Visual Inspection}\label{chap5.1}
\subsubsection{The Total Bar Fraction}\label{chap5.1.1}

\begin{figure*}[ht!]
\includegraphics[bb = 20 440 570 795, width = 0.95\linewidth, clip=]{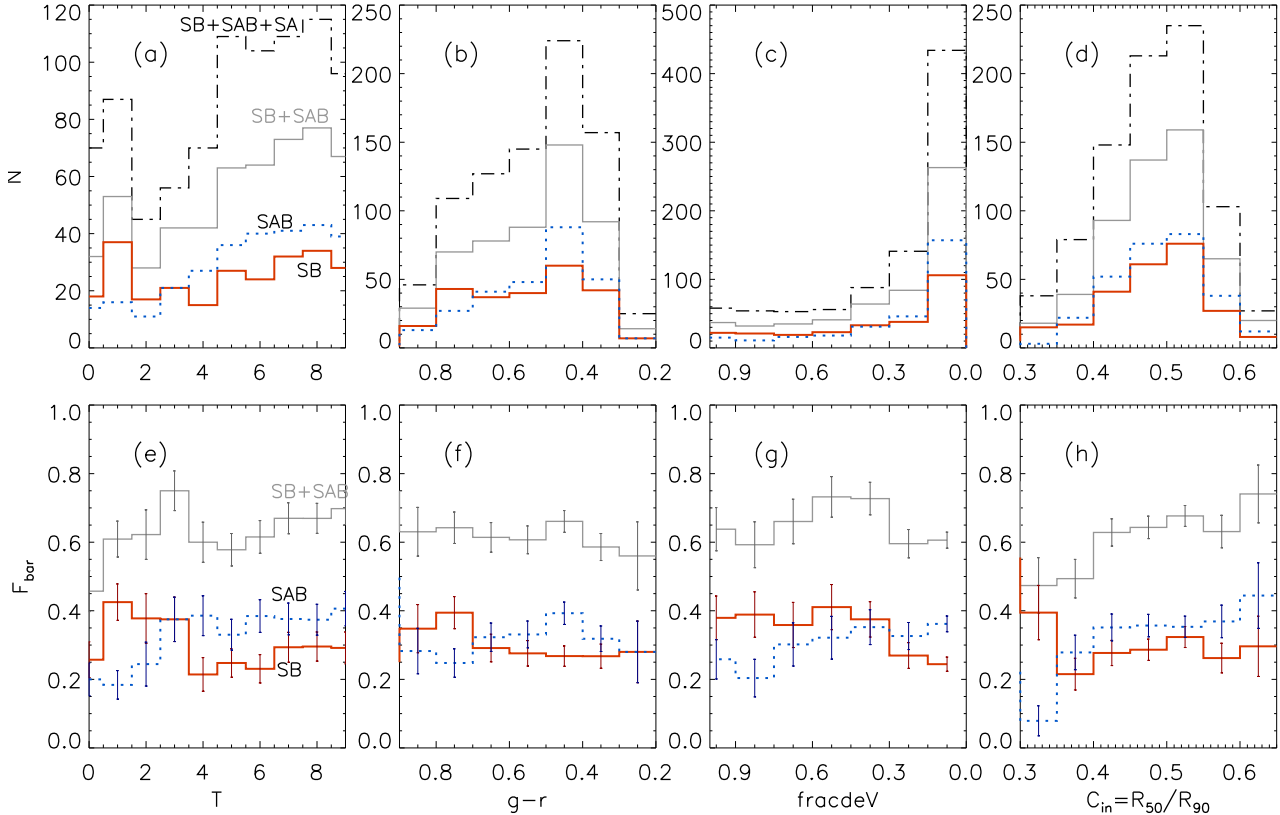}
\caption{Dependence of the number and fraction of barred galaxies from \citetalias{Ann15} catalog on (a) Hubble-sequence, (b) $\it g-r$, (c) fracdeV, and (d) $C_{\rm in}$. The top row shows the number of barred galaxies in each bin and the bottom row the bar fraction. The black dot-dashed line indicates the total galaxies and the gray solid line total barred galaxies. The red solid and blue dotted lines show SB and SAB galaxies. The x-axes of $\it g-r$ and fracdeV are converted from right to left, and the leftward of all panels presents the characteristics of early-type spirals. \label{fig5.1}}
\end{figure*}

We investigated the dependence of the fraction of barred galaxies classified by \citetalias{Ann15} on the properties of the host galaxies. In Fig. \ref{fig5.1}, the top row shows histograms of the number of barred galaxies and the bottom row the fractions of bars as a function of T, $(\it g-r)$, fracdeV, and $C_{\rm in}$ from left to right. The black dot-dashed and gray solid lines represent the total galaxies and the total barred galaxies, respectively. Strong bars (SB) and weak bars (SAB) are represented by red solid and blue dotted lines. The uncertainties are $[f(1-f)/N]^{1/2}$ for the fraction $f$ and number of galaxies $N$ in a given bin, representing statistical uncertainty in the fraction \citep{2008ApJ...675.1141S}. We reverse the x-axes of $(\it g-r)$ and fracdeV from right to left for easy comparison with other parameters. Accordingly, in all panels, the leftward corresponds to early-type spirals with low T, high $(\it g-r)$, high fracdeV, and low $C_{\rm in}$, while the rightward to late-type spirals which have large T, small $(\it g-r)$, small fracdeV, and high $C_{\rm in}$. We notice that our sample contains lots of galaxies with late-type (T $\geq$ 5), blue color (0.3 $\leq$ $\it g-r$ $\leq$ 0.5), small bulge (fracdeV $\leq$ 0.15), and less concentration (0.45 $\leq C_{\rm in} \leq$ 0.55) (Fig. \ref{fig5.1}a-d). 

The gray solid lines (bottom row of Fig. \ref{fig5.1}) that represent the bar fractions of both strong (SB) and weak bars (SAB) do not show any significant trend as a function of T, $(\it g-r)$, or fracdeV. They seem to be roughly constant. Previous studies showed similar results when both strong and weak bars are considered \citep{1999ASPC..187...72K, 2000AJ....119..536E, 2017ApJ...845...87L}. In more detail, the overall bar fractions seem to show big or small double peaks on T, $(\it g-r)$, and fracdeV. One peak appears in each panel around early-type spirals with T = 3, $(\it g-r)$ $\simeq$ 0.75, and fracdeV $\simeq$ 0.9. The other peak is located around late-type spirals or intermediate-type spirals with T $\geq$ 7, $(\it g-r)$ $\simeq$ 0.45 and fracdeV $\simeq$ 0.5. This is consistent with recent studies by visual inspection dealing with only strong or obvious bars \citep{2010ApJ...714L.260N, 2011MNRAS.411.2026M, 2012ApJ...745..125L}. They showed a more dramatic peak above $(\it g-r)$ $\simeq$ 0.7 and a small peak below $(\it g-r)$ $\simeq$ 0.4. We also see more conspicuous peaks around early-type spirals when we separate strong bars from weak bars (red solid lines in Fig. \ref{fig5.1}e-f). When it comes to the light concentration (Fig. \ref{fig5.1}h), the overall bar fraction obviously increases as host galaxies are less-concentrated, showing no double peaks. However, this monotonous trend is the result of the steady increase of weak bars with $C_{\rm in}$ (blue dotted line in Fig. \ref{fig5.1}h) and strong bars show the highest peak in the most concentrated bin (red solid line in Fig. \ref{fig5.1}h).

\subsubsection{Different Dependence of Strong and Weak Bar Fractions on Host Galaxy Properties} \label{chap5.1.2}

Therefore, we need to analyze the bar fractions of SB and SAB galaxies, separately. Counting SBs (red solid line) and SABs (blue dotted line) separately shows that $F_{\rm SB}$ and $F_{\rm SAB}$ have different distributions against the properties of host galaxies. The double peaks shown in the whole bar fractions are the combination of one peak by strong bars and another by weak bars: in Fig. \ref{fig5.1}e, strong bars are dominant in early-type spirals with T $\leq$ 3 while weak bars are prominent in intermediate- and late-type spirals with T $\geq$ 3. When it comes to the $(\it g-r)$ color (Fig. \ref{fig5.1}f), $F_{\rm SB}$ has a peak at red spirals with $(\it g-r)$ = 0.75 and $F_{\rm SAB}$ is peaked at blue spirals with $(\it g-r)$ = 0.45. As for fracdeV (Fig. \ref{fig5.1}g), strong bars are frequent in bulge-dominated systems with fracdeV $\geq$ 0.3 while weak bar fraction slowly increases as fracdeV decreases. We also found peaks of $F_{\rm SB}$ and $F_{\rm SAB}$ at different $C_{\rm in}$ in Fig. \ref{fig5.1}h: strong bars increase steeply at the most concentrated systems (lowest $C_{\rm in}$) whereas weak bars rise toward less-concentrated systems (higher $C_{\rm in}$) and have a peak at the least-concentrated system. In other words, strong and weak bars have their own peaks at different ranges. Strong bars are more frequent in early-type, red, bulge-dominated, and the most concentrated spirals while weak bars are frequent in intermediate-type and late-type, blue, disk-dominated, and less-concentrated spirals. The correlation between the bar types and host galaxy properties can solve some contradictory results seen in previous studies. We will discuss it more in \S\ref{chap6.1}. 

Our findings are consistent with lots of previous studies. Earlier study of \citet{1987JKAS...20...49A} showed that early-type spirals with higher spheroid-to-disk ratio have stronger bars. \citet{2005MNRAS.364..283E} have mentioned late-type spirals have weak bars twice as many as strong bars by analyzing the sample of \citet{1995AJ....109.2428M}. \citet{2000AJ....120.2835A} have shown that SBs and SAs increase toward early-type spirals and SABs are frequent in late-type spirals. \citet{2011MNRAS.411.2026M} also showed that the bar fraction increases with the value of fracdeV by using Galaxy Zoo project which mainly dealt with obvious bars, although it appeared to be more rapidly increasing than in our work. Similar dependences have been reported for flat and exponential bars: flat bars are frequent in early-type spirals while late-type spirals mainly have exponential bars \citep{1985ApJ...288..438E, 1989ApJ...342..677E, 1986PASP...98...56B, 1996AJ....111.2233E, 2011MNRAS.415.3627H}. When it comes to the bar length, early-type or red spirals have longer bars than late-type or blue spirals \citep{2005MNRAS.364..283E, 2012ApJ...745..125L}. Early-type spirals showed higher bar strength estimated by the bar luminosity \citep{1987JKAS...20...49A} and by the Fourier amplitude \citep{1990ApJ...357...71O, 2009ApJ...692L..34L}.

We noticed another interesting feature in Fig. \ref{fig5.1}e. The bar fraction declines abruptly at S0/a galaxies (T=0) despite the fact that not only early-type spirals have a high fraction of strong bars but also bulge-dominated and highly concentrated systems exhibit a high fraction of strong bars. Although it is not exactly about S0/a, we often find previous studies that showed lack of bar in S0 galaxies compared to spirals \citep{Agu09, 2009ApJ...692L..34L, 2015ApJS..217...32B, 2017ApJ...845...87L}. \citet{2009ApJ...692L..34L} mentioned that S0 galaxies have higher fraction of ovals/lenses, which might have been bars if not weakened by central concentration. We suspect another possibility that lack of gas in S0/a and S0 galaxies makes it hard to drive bar instability if the formation of bar is delayed until gas is removed from these galaxies.

\subsection{Different Tendencies in Bar Fraction by Automatic Classification}\label{chap5.2}

\begin{figure*}[hbtp]
\includegraphics[bb = 20 600 570 790, width = 0.95\linewidth, clip = ]{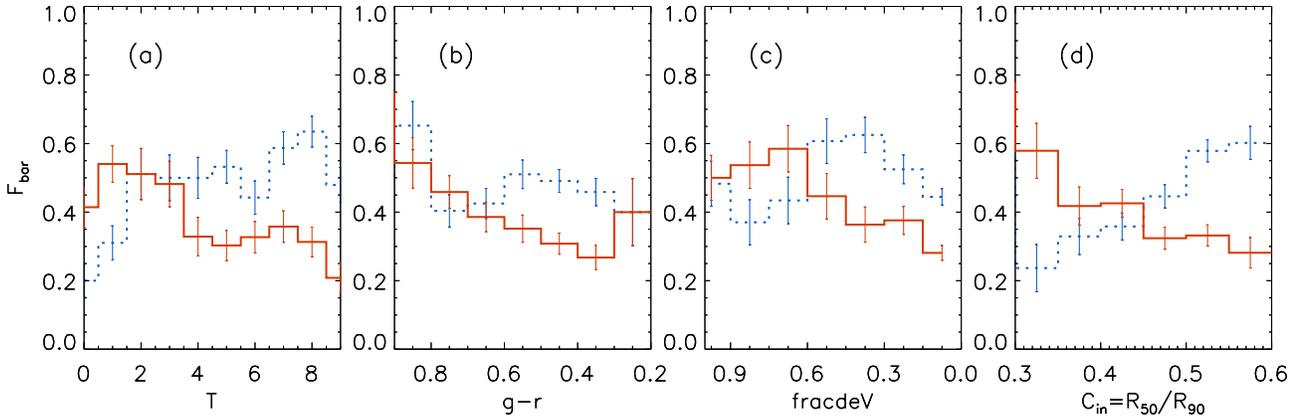}
\caption{Dependence of the bar fraction by ellipse fitting method (blue dotted lines) with $g_{\rm obs}$+5$^\circ$+0.25 and Fourier analysis (red solid lines) with \citetalias{Lau02} criteria on (a) Hubble sequence, (b) $(\it g-r)$ color, (c) fracdeV, and (d) $C_{\rm in}$. The leftward of all panels represent the properties which early-type spirals have.\label{fig5.2}}
\end{figure*}

We display the dependence of bar fraction on host galaxy properties by using ellipse fitting method (blue dotted lines) and Fourier analysis (red solid lines) in Fig. \ref{fig5.2}, in the same manner as in Fig. \ref{fig5.1} except for the method to select barred galaxies. We chose the condition of $g_{\rm obs}+5^\circ+0.25$ for ellipse fitting and the criterion of \citetalias{Lau02} for Fourier analysis in order to compare with previous studies. These are typical conditions used in previous studies and yield good agreements with visual classification (refer \S \ref{chap3}). Weak bars are not treated separately here because the fraction is too small in ellipse fitting method (Table\ref{Table4}) and not defined in \citetalias{Lau02}. 

At a glance, the distributions of bar fraction seem to be quite different depending on the method to select bars. Fourier analysis shows higher bar fractions toward early-type spirals, which increase as the host galaxies are red, bulge-dominated, and more concentrated (red solid lines in Fig. \ref{fig5.2}a-d). These tendencies are consistent with previous studies by Fourier analysis \citep{Agu09, 2009ApJ...692L..34L}. And the bar fractions from these studies resemble the distribution of $F_{\rm SB}$ by visual inspection, although not exactly the same. The consistency can be understood because the criterion of \citetalias{Lau02} mainly detects strong bars. Actually, the overall bar fraction by \citetalias{Lau02} just stays around 36\%, which is slightly higher than the frequency of visually detected strong bars. We, however, have to be cautious that the errors of the Fourier method is particularly large in early-type spirals with a large bulge. Fortunately, the errors do not seem to significantly influence the distribution of bar fraction as a function of Hubble type because the Fourier method sometimes misses barred galaxies and other times determines non-barred galaxies as barred galaxies in early-type spirals as discussed in \S\ref{chap3.3}.

On the contrary, the bar fractions yielded by the ellipse fitting method increase toward late-type spirals, i.e., in disk-dominated and less-concentrated systems. These tendencies agree with previous studies obtained by the ellipse fitting method \citep{2008ApJ...675.1194B, 2009A&A...497..713B, Agu09} and resemble the distribution of $F_{\rm SAB}$ by visual inspection in some ways. But these look opposite from those by Fourier analysis. Consequently, though we reproduced the bar fraction as a function of galaxy properties consistent with previous studies by using ellipse fitting and Fourier analysis, we faced a contradiction that the distribution of bar fractions are different even for the same sample. \citet{Agu09} also compared the two automatic classifications and found that the bar fraction by Fourier analysis is lower in late-type spirals than that by ellipse fitting. They reported that Fourier analysis is less efficient in detecting bars of the late-type spirals with lenses or strong spiral arms. Our analysis is slightly different from their findings. We suspect two effects as causes for the discrepancy. One is that two methods accept weak bars of different fraction as their barred galaxies: the ellipse fitting method classifies more visual SABs as barred galaxies compared to Fourier analysis. Secondly, the systematic errors of ellipse fitting technique which could miss bars mainly in early-type spirals with a large bulge could influence the bar fraction as a function of galaxy properties as expected in \S\ref{chap3.2}. 

\section{DISCUSSION}\label{chap6}

\subsection{Bar Fraction as a Function of Host Galaxy Properties}\label{chap6.1}

This work is partly motivated by the contradictory bar fraction results as a function of host galaxy properties shown in previous studies. We find a trend that studies with relatively higher bar fractions between 45\%$\sim$66\% show frequent bars in late-type spirals  \citep{2008ApJ...675.1194B, Agu09, 2009ApJ...696..411W, 2015ApJS..217...32B, 2015MNRAS.446.3749Y, 2018Erwin} while studies with lower bar fractions around 30\% report abundant bars in early-type spirals \citep{2008ApJ...675.1141S, Agu09, 2011MNRAS.411.2026M, 2012ApJ...745..125L, 2012ApJS..198....4O, 2013ApJ...779..162C, 2015A&A...580A.116G, 2016AA...595A..67C}. In the previous section, we showed the different dependence of bar fraction on the Hubble type for strong versus weak bars. So we can understand that studies including more weak bars would show increasing bar fraction toward late-type spirals whereas the studies that mainly deal with strong bars would show prominent bar fraction in early-type spirals.

In fact, \citet{2018Erwin} has also questioned the different Hubble-type dependence for SBs and SABs. However, he could not find significantly distinct characteristics for $(\it g-r)$ and the stellar mass $M_\ast$, and has shown only the diverging bar fractions of SB and SAB for very high gas mass ratios. Our analysis also confirms that the difference is more prominent for the Hubble type and bulge dominance rather than for $(\it g-r)$ and $M_\ast$ in Fig. \ref{fig6.1} and Fig. \ref{fig6.3}. This suggests that $(\it g-r)$ and $M_\ast$ have little effect on the different formation or evolution of strong versus weak bars, although they are important parameters for bar formation and evolution. 

\citet{2010ApJ...714L.260N} understood the discrepancy was caused by different mass range of sample galaxies they used in previous studies \citep{2008ApJ...675.1141S, 2008ApJ...675.1194B}. They showed bimodal distribution of bar fraction as a function of host galaxy properties from visually obvious bar fraction of $\sim$30\%. \citet{2011MNRAS.411.2026M, 2012Masters} and \citet{2012ApJ...745..125L} have found similar trends of bar fraction with a high peak at red color and a very small peak at blue color from visually obvious bars which constitutes $\sim$30\% of the spiral galaxies. They argued that the disagreement was caused by excluding red disk galaxies using color cut in the study of \citet{2008ApJ...675.1194B}. 

\begin{figure*}[htbp]
\includegraphics[bb = 20 458 570 800, width = 0.99\linewidth, clip=]{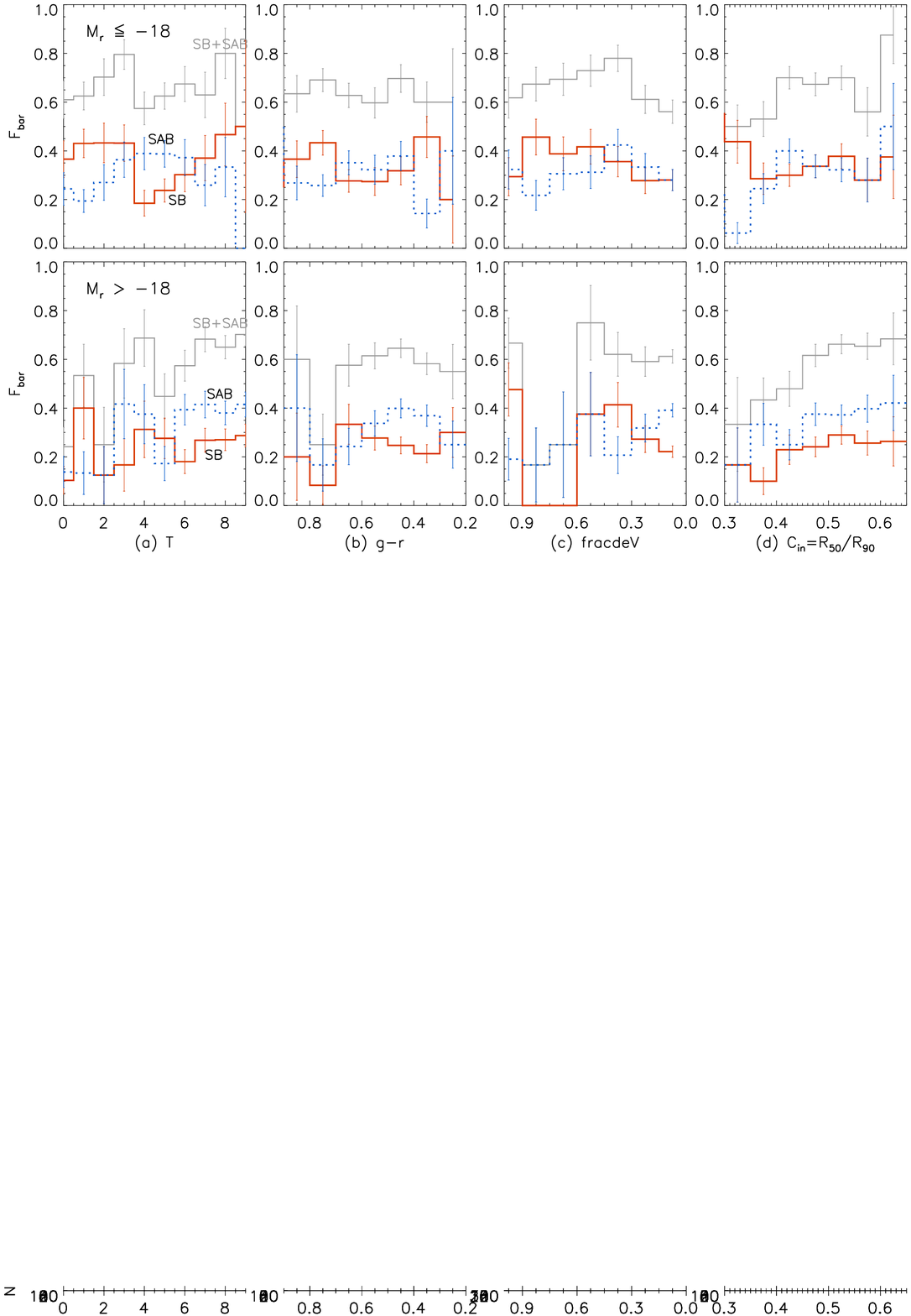} 
\caption{Dependence of the fraction of barred galaxies from \citetalias{Ann15} catalog on (a) Hubble-sequence, (b) $g-r$,  (c) fracdeV, and (d) $C_{in}$. The top row shows the subsample of brighter galaxies with $M_r \leq -18$ and bottom row fainter galaxies with $M_r > -18$. We present total barred galaxies (gray solid line), strongly barred galaxies (red solid line), and weakly barred galaxies (blue dotted line), respectively. \label{fig6.1a}}
\end{figure*}

In order to test the possible sample bias, we split our sample into two groups, bright ($M_r \le -18$) versus faint ($M_r > -18$) galaxies in Fig. \ref{fig6.1a}. They show a roughly similar trend with the total sample: roughly equal SBs in early- and late-type, and more SABs in late-type spirals. However, we note that another high peak of SBs appears at late-type and blue spirals when faint galaxies are excluded. It can explain the higher bar fraction at low-mass galaxies shown in \citet{2010ApJ...714L.260N}, as discussed by \citet{2015A&A...580A.116G}. 

In addition, we emphasize that the method to classify barred galaxies causes the different bar fraction on the host galaxy properties. It is because not only they contain different fraction of weak bars but also they often misclassify bulge-dominated galaxies. Besides, \citet{2018Erwin} suspected the poor spatial resolution of SDSS as the reason of discrepancy between previous studies. \citet{2013ApJ...779..162C} provided another view that the mass-dependence of bar fraction again depends on the specific star formation rate (SSFR). 

Lastly, we investigate the bar fraction in terms of the stellar mass. In Fig. \ref{fig6.1}, we display the bar fraction as a function of mass by visual inspection and automatic classification. Similarly with other properties, the fraction of strong bar by visual inspection increases with mass (red solid line in Fig. \ref{fig6.1}a). This result is consistent with previous studies mainly dealing with obvious bars by visual inspection \citep{2012Masters, 2014MNRAS.438.2882M, 2015A&A...580A.116G}. On the other hand, weak bars are hardly influenced by mass (blue dotted line in Fig. \ref{fig6.1}a). For the automatic classification, both methods yield strong trends that increase with mass (Fig. \ref{fig6.1}b). Although this seems different from the tendency in other properties shown in \S \ref{chap5.2}, it agrees with previous studies by automatic classification \citep{2015A&A...580A.116G, 2016AA...595A..67C}. When it comes to mass, we do not find the opposite tendency between the Fourier analysis and the ellipse fitting method. However, we emphasize that the estimate of stellar mass for our galaxy sample may have significant errors due to their peculiar velocities. Likewise, we need to be cautious about the results related to the galaxy mass in previous studies that deal with nearest galaxies \citep{2016A&A...587A.160D, 2018Erwin}. 

\begin{figure}[htbp]
\includegraphics[bb = 75 440 380 620, width = 0.99\linewidth, clip=]{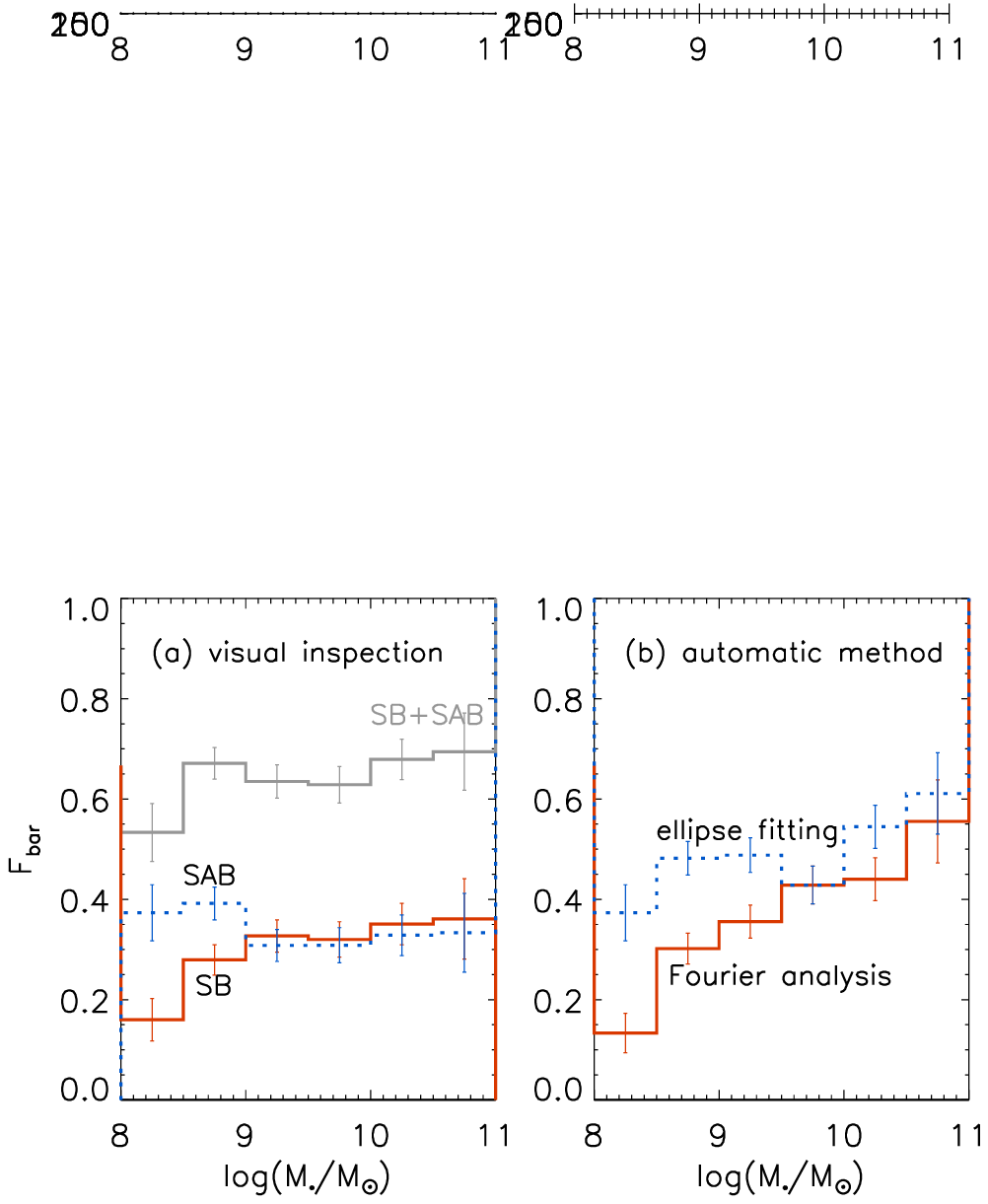} 
\caption{Bar fraction as a function of the stellar mass (a) by visual inspection and (b) by automatic classification. \label{fig6.1}}
\end{figure}

\subsection{Properties of Bars: SBs versus SABs}\label{chap6.2}

In general, earlier classification by visual inspection showed SBs of 30\% and SABs of 30\% in nearby spiral galaxies \citep{1973ugcg.book.....N, 1987rsac.book.....S, RC3}. More recently, \citetalias{Ann15} and \citet{2015ApJS..217...32B} also presented similar results. They usually inspect the shape, relative bar length and contrast to distinguish SABs from SBs. 

We find a similar dichotomy of bars in previous studies \citep{1985ApJ...288..438E, 1989ApJ...342..677E, 1986PASP...98...56B, 1996AJ....111.2233E, 1997AJ....114..965R, 2015ApJ...799...99K}. They investigated two types of bars: flat and exponential profiles in surface brightness. Flat bars have nearly constant light distributions along the bar whereas those of exponential bar decrease exponentially. They also differ in their structures and the intensity contrast between the bar and the disk: flat bars are longer, wider, and stronger than exponential bars, and have higher contrast compared to exponential bars. Besides, \citet{1992MNRAS.259..345A} explained that flat bars could have roughly rectangular orbits around the end of bar through the stellar orbit calculation. These properties can also be explained by the locations of resonances \citep{1979MNRAS.187..101L, 1980A&A....92...33C, 1981A&A....99..362S, 1989ApJ...343..608C, 1992MNRAS.259..345A, 2002MNRAS.333..861S}. Flat density profile develops in crowding stellar orbits between the inner 4:1 resonance and corotation radius \citep{1993A&A...271..391C, 1985ApJ...288..438E, 1996AJ....111.2233E}. Exponential bars end near the inner Lindblad resonance and do not have such crowding orbits \citep{1979MNRAS.187..101L, 1985ApJ...288..438E, 1996AJ....111.2233E}. Therefore, flat bars and exponential bars may be expected to have different pattern speeds based on the value of $R = R_{cr}/R_{bar}$ where $R_{cr}$ and $R_{bar}$ are the radius of the corotation resonance and the bar, respectively \citep{2000ApJ...543..704D, 2003MNRAS.345..406V}. Although some studies reported the observational lack of slow bars \citep{2000ApJ...543..704D, 2015A&A...576A.102A}, others showed that the pattern speed of bars roughly depends on the Hubble type: fast bars in early-type spirals and slow bars in late-type spirals \citep{1998AJ....116.2136A, 2008MNRAS.388.1803R}. More observational data will help understand the relation between the density profile and the pattern speed of bars.

\begin{figure}[htbp]
\includegraphics[bb = 20 420 410 770, width = 0.99\linewidth, clip=]{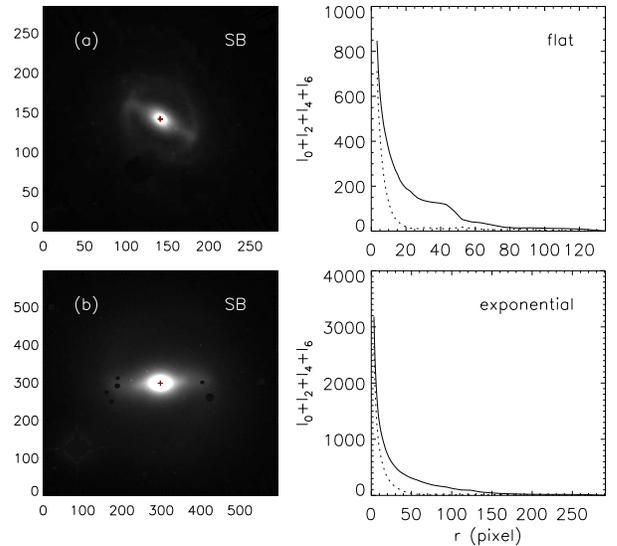} 
\caption{The examples for a flat bar and an exponential bar. The left panels show images and the right panels the bar intensity and inter-bar intensity profiles. The bar intensity (solid line) is calculated by $I_0+I_2+I_4+I_6$ from Fourier analysis and the inter-bar intensity (dotted line) by $I_0-I_2+I_4-I_6$. (a) SB and flat bar. The flat bar has a plateau in the bar intensity profile. (b) SB and exponential bar. \label{fig6.2}}
\end{figure}

Therefore, the shape, length, strength, and pattern speed of bars seem to be related to the dichotomy of flat and exponential bars. We investigated whether our SBs and SABs defined by their shape, rectangular or oval, are related to flat or exponential bars. We classified our sample galaxies into flat or exponential profiles. We investigated the bar intensity profiles of $I_{\rm 0}+I_{\rm 2}+I_{\rm 4}+I_{\rm 6}$ calculated in \S\ref{chap2.2.2} \citep{1990ApJ...357...71O, 2000AA...361..841A} and classified galaxies which show a plateau in the bar intensity profile as flat bars \citep{1985ApJ...288..438E, 1996AJ....111.2233E, 2015ApJ...799...99K} as shown in Fig. \ref{fig6.2}. We described the percentage of flat and exponential profiles for each class, SB, SAB, and SA, in Table \ref{Table5}. Even non-barred galaxies have flat profiles. Flat profiles can be formed not only by bar structures but also by rings, winding spirals, and bright clusters around the center. Nevertheless, we confirm that flat profiles increase toward strong bars and exponential profiles increase toward non-barred galaxies. Although there is no direct match between the bar type (SBs or SABs) and the bar luminosity profile (flat or exponential), flat bars are more dominant in SBs than in SABs.

\begin{deluxetable}{l|ccc}
\tablecaption{Classification for surface brightness profile of $I_{\rm 0}+I_{\rm 2}+I_{\rm 4}+I_{\rm 6}$ \label{Table5}}
\tablehead{
 \colhead{profile classification} &
 \colhead{SB} &
 \colhead{SAB} &
 \colhead{SA} \\
}
\startdata
flat        & 63.6\% & 43.6\% & 38.2\% \\
exponential & 36.4\% & 56.4\% & 61.8\% \\
\enddata
\end{deluxetable}

\subsection{Bar Formation and Evolution in Early-type and Late-type Spirals}\label{chap6.3}

\begin{figure*}[htbp]
\includegraphics[bb = 0 570 590 770, width = 0.98\linewidth, clip=]{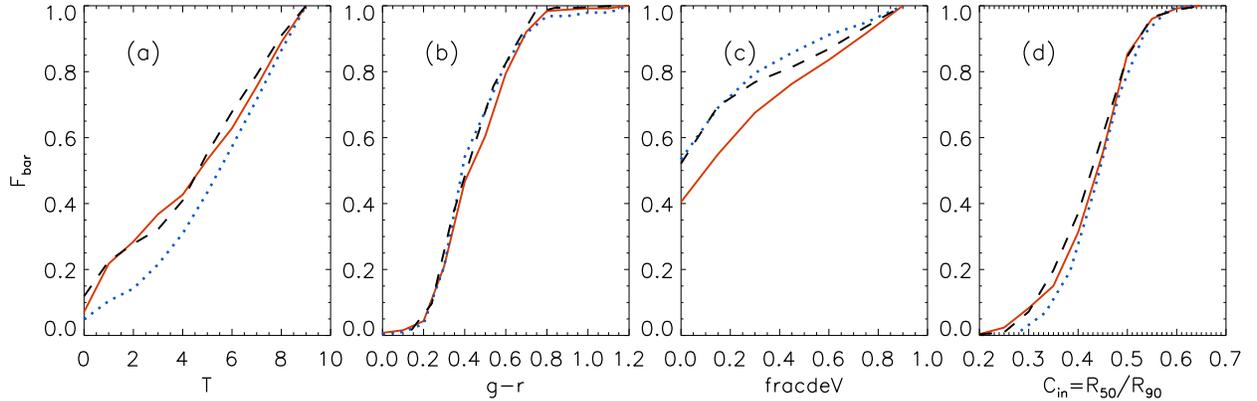} 
\caption{The cumulative distribution of $F_{\rm SB}$, $F_{\rm SAB}$ and $F_{\rm SA}$ as a function of T (a), $(\it g-r)$ color (b), fracdeV (c) and $C_{\rm in}$ (d). Solid lines and dotted lines represent SB and SAB, respectively and dashed lines SA. \label{fig6.3}}
\end{figure*}

We carried out the Kolmogorov-Smirnov (K-S) test to estimate how SBs and SABs are differently distributed with respect to the host galaxy properties. Table \ref{Table3} presents the probability of K-S test between bar families. Small probability values mean that two distributions are statistically different. The cumulative distributions of histograms for $F_{\rm SB}$ (Solid line), $F_{\rm SAB}$ (dotted line), and $F_{\rm SA}$ (dashed line) are shown in Fig. \ref{fig6.3}. We find that SBs have totally different distributions from those of SABs as a function of the Hubble sequence as shown in Fig. \ref{fig5.1}e. This is consistent with the analysis of \citet{2000AJ....120.2835A}. They argued that SABs would be the expanded form of non-barred galaxies in late-type spirals. 

On the other hand, the parameter that distinguishes SB from both SA and SAB is the bulge property represented by fracdeV. SABs have similar characteristics with SAs when it comes to fracdeV but are distinguished from SAs for $C_{in}$. We can not find any difference between bar families in $(\it g-r)$. Similar characteristics have appeared for the mean values of T, $(\it g-r)$, fracdeV, and $C_{\rm in}$ for SA, SAB, and SB galaxies (Table \ref{Table7}). We find that SB galaxies have higher mean value of fracdeV than SAs and SABs, whereas SAB galaxies have higher T compared to SAs and SBs. Although we do not find significant differences in $(\it g-r)$  and $C_{\rm in}$, SAs have slightly lower mean values in $(\it g-r)$ and SABs have slightly higher in $C_{\rm in}$.  

Accordingly, we surmise that strong and weak bars prefer different inner galactic structures for their formation and evolution. They would experience different processes in early-type and late-type spirals. Especially, we think that it is the bulge properties that make strong bars different from weak bars. These different bar properties in early- and late-type spirals have been reported similarly in previous studies: flat bars and buckled bars are dominant in early-type spirals whereas exponential bars and unbuckled bars are mainly found in late-type spirals \citep{1985ApJ...288..438E, 1989ApJ...342..677E, 1986PASP...98...56B, 1996AJ....111.2233E, 2017MNRAS.468.2058E, 2017ApJ...845...87L}. Different characteristics of bars seem to be closely related to the difference in the structure of galaxies. 

\begin{deluxetable}{l|ccc}
\tablecaption{The probabilities from K-S test for bar fraction as a function of host galaxy properties between SB, SAB and SA\label{Table3}}
\tablehead{
  \colhead{properties}  & 
  \colhead{SB vs. SAB} & 
  \colhead{SAB vs. SA} &
  \colhead{SB vs. SA} \\
}
\startdata
T           & 0.0018 & 0.0067 & 0.7138 \\
$\it g-r$   & 0.2332 & 0.7188 & 0.2997 \\
fracdeV     & 0.0008 & 0.6758 & 0.0003 \\
$C_{in}$    & 0.0948 & 0.0345 & 0.6445 \\
\enddata
\end{deluxetable}

\begin{deluxetable}{l|cccc}[hbp]
\tablecaption{Mean values for properties of SA, SAB, SB galaxies\label{Table7}}
\tablehead{
  \colhead{Properties}  & 
  \colhead{T} & 
  \colhead{$\it g-r$} &
  \colhead{fracdeV} &
  \colhead{$C_{\rm in}$} \\
}
\startdata
SA & 4.7 & 0.53& 0.26 & 0.47 \\
SAB & 5.6 & 0.57 & 0.24 & 0.49 \\
SB  & 4.8 & 0.56 & 0.34 & 0.47 \\
\enddata
\end{deluxetable}

Theoretical studies and numerical simulations have shown that bars form, grow, and weaken via active interactions with other components of host galaxies such as bulge, disk, halo, and gas component. For early-type spirals, their spheroidal components stabilize the disk and delay the formation of bars \citep{1964ApJ...139.1217T, 1973ApJ...186..467O, 1981A&A....96..164C, 1986MNRAS.221..213A}. However, their bar grows gradually by interaction with a bulge and a halo which gain angular momentum from the disk and exert a dynamical friction on the bar \citep{1980A&A....89..296S, 1985MNRAS.213..451W, 2002ApJ...569L..83A, 2003MNRAS.341.1179A}. Consequentially, the bar pattern speed continuously slows down and the corotation radius moves toward outer regions, driving more particles to be trapped \citep{1985MNRAS.213..451W, 1991MNRAS.250..161L, 1992ApJ...400...80H, 1996ASPC...91..309A, 1998ApJ...493L...5D, 2003MNRAS.341.1179A}. \citet{1993A&A...271..391C} explained that early-type spirals with large bulge-to-disk ratios have a large maximum in the distribution of $\Omega-\kappa/2$ and a large pattern speed initially. Therefore, the corotation radius is located at a small radius and bars can transfer the angular momentum to stars near the corotation and outer resonance regions. Enough interactions make the pattern speed slow down and intersect $\Omega-\kappa/2$. Eventually, inner Lindblad resonances occur and it is possible to develop flat bars and ring structures. 

On the other hand, in the case of late-type spirals, although they readily form bars due to less-massive halo, bars can not grow because of few interacting stars \citep{1981A&A....96..164C, 1993A&A...271..391C, 2002MNRAS.330...35A, 2002ApJ...569L..83A}. They have a low $\Omega-\kappa/2$ and pattern speed. Thereby, the corotation radius resides far beyond the end of the bar. In this case, the disk scale length determines the bar length and the bar has an exponential density profile \citep{1993A&A...271..391C}. 

Bars in early-type and late-type spirals also showed different results in their destruction by central mass concentration (CMC) such as a supermassive black hole, central disk, dense central stellar clusters, and bar-driven gas. First, bars in disk-dominated galaxies are totally destroyed while bars in massive halo systems just weakens in their strength with respect to the same CMC \citep{2005MNRAS.363..496A}, due to the extra angular momentum absorbed by the halo. Secondly, late-type spirals have abundant gas, and bars in late-type spirals seem to be more easily weakened. It is not only because the gas inflow can give angular momentum to bars but also because the gas inflow largely contributes to the massive CMC even when they have the same halos \citep{2007ApJ...666..189B, 2013MNRAS.429.1949A}. Finally, disk thickness also drives the difference in bars. The disk scale height is an indicator of the phase-space density in the central region. Thin disk with high phase-space density and smaller random velocity stimulates the buildup of the CMC. On the other hand, bars in thick disks with larger random velocity lose more angular momentum and rotate slower in a similar way as the bulge \citep{2009MNRAS.398.1027K}.

Consequentially, bars in late-type spirals are not only easily generated and but also easily destroyed with rare growth. However, bars in early-type spirals gradually grow by interacting with their bulge, thick disk and halo, and are not easily destroyed. Therefore, bars in early-type spirals would be more stable to survive, although they need longer time to form compared to bars in late-type spirals. Nevertheless, we seem to find similar bar fractions of SBs and SABs in the local universe. We speculate that this is caused by the combination of frequent formation and short lifetime of weak bars and rare formation and longer lifetime of strong bars.

\subsection{Bar Growth and Destruction by Bulge and CMC}\label{chap6.4}

The growth and destruction of bars seem to depend on two competitive effects between bar growth by the bulge or halo and bar destruction by the CMC. We have two parameters, fracdeV and $C_{\rm in}$, which can roughly estimate the bulge dominance and CMCs, respectively. Although both of them have often used as the indicator of the bulge size \citep{1996MNRAS.279L..47A, 2000AJ....120.2835A, 2005ApJ...635L..29P, 2011MNRAS.411.2026M}, we have shown that not only the correlation between two parameters is relatively weak (Fig. \ref{fig5}f) but also the dependence of bar fraction on two parameters appear to be different (Fig. \ref{fig5.1}g and h). If we consider that fracdeV obtained by profile model fitting would better reflect the bulge properties while $C_{\rm in}$ calculated by comparing $R_{\rm 50}$ and $R_{\rm 90}$ could be representative for the CMC due to the SMBH, bar-driven gas and bar itself as well as the bulge, we can obtain the observational hint for the effect of the bulge and CMC on the evolution of bars. The parameter $C_{\rm in}$ has been explained as the indicator of the CMC in \citet{2010ApJ...714L.260N} and \citet{2012ApJ...745..125L}.

First look on Fig. \ref{fig5.1}g shows that the distributions of $F_{\rm SB}$ and $F_{\rm SAB}$ have the opposite slope with respect to the fracdeV: strong bars are abundant in bulge-dominated spirals and weak bars gradually increase toward the spirals with small bulges. It seems hard for bars to grow stronger in galaxies with a small bulge. This confirms the simulations that bars grow stronger by interaction with bulges which deprive of angular momentum from bars \citep{2002MNRAS.330...35A, 2002ApJ...569L..83A, 2003MNRAS.341.1179A}.

The dependence of $F_{\rm SB}$ and $F_{\rm SAB}$ on $C_{\rm in}$ seems to be more complex as seen in Fig. \ref{fig5.1}h. Strong bars and weak bars have a maximum in bar fraction on the opposite end bins: $F_{\rm SB}$ is peaked at the most concentrated spirals, whereas $F_{\rm SAB}$ at the least concentrated spirals. If the most concentrated $C_{\rm in}$ bin is excluded, both $F_{SB}$ and $F_{SAB}$ are slightly increasing toward less-concentrated spirals. However, the dependence of $F_{\rm SB}$ and $F_{\rm SAB}$ look different on the fracdeV. It seems that the light concentration doesn't help all types of bars to form or survive. This observation is consistent with the theoretical expectation that the CMCs weaken or completely destroy bars \citep{1990ApJ...363..391P, 1990ApJ...361...69H, 1993ApJ...409...91H, 1996ApJ...462..114N}.  Although more recent simulations showed that bars are robust structures, and the current masses of SMBHs and diffuse gas concentration are inefficient in destroying bars totally, they all found that the growth of CMCs decreases the bar strength \citep{2004ApJ...604..614S, 2005MNRAS.363..496A, 2006ApJ...645..209D}. Our observation also suggests that the bar growth still seems to be regulated by the CMC.

On the other hand, the distributions of $F_{\rm SB}$ and $F_{\rm SAB}$ on $C_{\rm in}$ show some difference in the most concentrated bin (Fig. \ref{fig5.1}h). In the most concentrated systems, the fraction of strong bars abruptly increases to a high ratio, whereas that of weak bars greatly decreases. Therefore, the distribution of $F_{\rm SB}$ appears bimodal (Fig. \ref{fig5.1}h), which is consistent with \citet{2010ApJ...714L.260N} who dealt with strong bars. We find a hint to understand the unexpected increase of strong bars in the most concentrated systems from the correlation between fracdeV and $C_{\rm in}$ in Fig. \ref{fig5}. All of the most concentrated spirals with $C_{\rm in} \leq 0.3$ are distributed in the region with a massive bulge of fracdeV $\geq$ 0.6. Therefore, we may conclude that the bar growth by bulge is more prevalent compared to the weakening of bars by CMC in the most concentrated galaxies. It agrees with the simulation of \citet{2005MNRAS.363..496A} which showed that the effect of CMCs becomes weak in the bulge-dominated systems. 

\begin{figure}[htbp]
\includegraphics[bb = 70 400 550 750, width = 0.98\linewidth, clip=]{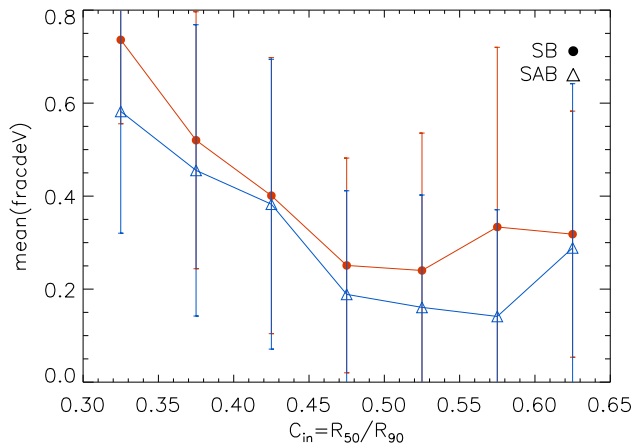} 
\caption{The mean fracdeV of SBs and SABs as a function of $C_{\rm in}$. The filled circles and open triangles indicate strong bars and weak bars, respectively. \label{fig6.4}}
\end{figure}
In Fig. \ref{fig6.4}, we compare the mean values of fracdeV of SBs and SABs for each $C_{\rm in}$. Although the distribution of fracdeV in the same range of $C_{\rm in}$ shows large scatters, we find two dominant features. Firstly, as the system is more-concentrated, its mean fracdeV largely increases. Although it is partly due to the effect of weak correlation between two parameters, we can speculate that galaxies need more massive bulges for the bar survival when galaxies are more concentrated. Secondly, SB galaxies always have higher mean values of fracdeV compared to SABs for the same $C_{\rm in}$ over all range of $C_{\rm in}$. This supports the view that galaxies need more developed bulge for strong bars compared to weak bars for a given CMC. Therefore, we confirm that bars are formed and supported by the balance between the bulge and CMC. 

On the other hand, we find the increasing bulge ratio of strong bars again at $C_{\rm in} > 0.55$ despite their low CMCs in Fig. \ref{fig6.4}. There seems to be the difficulties for bars to grow stronger in galaxies with a very low concentration and need more prominent bulges for strong bars. It would explain why the $F_{\rm SB}$ does not increase any more when $C_{\rm in} > 0.55$ in Fig. \ref{fig5.1}h.

Lastly, we want to discuss the bulge component. When the value of fracdeV is large, we expect the existence of classical bulge and for small fracdeV galaxies with pseudo-bulges or no bulges. Therefore, Fig. \ref{fig5.1}g shows that strong bars are frequent in galaxies with classical bulges while weak bars in galaxies with pseudo-bulges. In general, pseudo-bulges have been known to be formed as a result of the secular evolution by bar. The bar-driven gas can broaden the vertical resonance, and it is believed to form pseudo-bulges which are rotationally supported with metal-rich and young stellar population \citep{1982ApJ...257...75K, 1990ApJ...363..391P, 1993IAUS..153..209K, 2004ARA&A..42..603K}. They are abundant in late-type spirals, although early-type spirals also have pseudo-bulges \citep{1994MNRAS.267..283A, 1995MNRAS.275..874A, 2002AJ....123..159C, 2004ARA&A..42..603K}. Studies that showed the increasing bar fraction toward late-type spirals argued that the frequent pseudo-bulges in late-type spirals would be related to the high bar fraction in late-type spirals \citep{2008ApJ...675.1194B}. However, we want to consider the fact that oval structures such as weak bars are enough to drive gas inflow and pseudo-bulge formation \citep{2004ARA&A..42..603K, 2018MNRAS.tmp.1396K} and the larger gas fraction helps grow CMC \citep{2007ApJ...666..189B, 2013MNRAS.429.1949A}. Therefore, frequent weak bars and abundant gas in late-type spirals seem to explain the frequent pseudo-bulges seen in late-type spirals well. 

\section{CONCLUSIONS}\label{chap7}

We have studied the bar fraction and the relation between the fraction and the properties of their host galaxies for the selected 884 spiral galaxies from a parent sample of 1,876 spiral galaxies of the \citetalias{Ann15} catalog. It is a volume-limited sample with $z < 0.01$ from SDSS/DR7, brighter than $M_{\rm r}$ = -15.2 and an inclination $i < 60^\circ$. Specifically, we have compared the result from each bar classification method, visual inspection (\citetalias{RC3} and \citetalias{Ann15} catalog), ellipse fitting, and Fourier analysis, in order to understand the bias caused from different methods. 

\begin{enumerate}
\item We confirm the consistency of around $\sim$80\% and inconsistency below 10\% between two independent visual inspection, \citetalias{RC3} and \citetalias{Ann15}. The agreement of ellipse fitting method with the visual inspection reaches up to 70\% $\sim$ 80\% in SA and SB classes, and we obtain the best agreement when we apply the PA transition ($\Delta \rm PA_{\rm tra}$) of $5^\circ$ to $g-$band deprojected images. However, this method misses about 15\% of visually strong barred galaxies. We note that it is caused by the large bulge in early-type spirals which hides the transition between a bar and a disk. Therefore, the classification based on the ellipse fitting method produces a systematic effect on the bar fraction as a function of Hubble sequence. We can get the highest consistency around 80\% with the visual inspection when we apply the combined criterion of the relative Fourier amplitude and constant phase suggested by \citetalias{Lau02}. However, it also often makes mistakes in classification, in particular, in early-type spirals because of the large bulge. Large bulges have given variations on the phase or inclined bulges have been confused as bar structures.   

\item We obtain different bar fractions from the same sample galaxies when we apply different classification methods, criteria, or conditions. The visual inspection by \citetalias{Ann15} shows the bar fraction of 30\% of SB and 33\% of SAB. The ellipse fitting method, in general, yields a bar fraction of over 48\%. The resultant bar fraction contains almost all visually determined SBs and half of visually determined SABs. Fourier analysis method with the criteria of \citetalias{Lau02} yields a lower bar fraction of 36\%. It mainly finds strongly barred galaxies. Therefore, the range of barred galaxies can be different depending on the methods to classify bars. In addition, even when we use the same method, different criteria and observing  wavelength also influence the bar fraction.  

\item The dependence on the host galaxy properties of bar fraction depends on bar types, strong and weak. The visual inspection yields a different correlation between bar types and host galaxy properties. Strong bars are more frequent in early-type, red, bulge-dominated, and most concentrated galaxies, while the fraction of weak bars increases toward the late-type, blue, disk-dominant, and less-concentrated. We propose that strong and weak bars have experienced different processes for their formation and evolution within different type of galaxies, early- and late-type spirals. Their similar bar fractions of $\sim$30\% are likely to be the result of the combination of frequent formation and dissipation of weak bars and rare formation and longer survival of strong bars.  

\item The dependence on the host galaxy properties of bar fraction depends on the methods to select bars. For the same sample, we obtain the opposite dependence on the host galaxy properties when we used different classification methods. Bars classified by the ellipse fitting method are frequent in late-type spirals, which resemble those of weak bars by visual inspection. On the contrary, fraction of bars identified by Fourier analysis increases toward early-type spirals and they are similar to those of strong bars by visual inspection. We suspect that it is due to the fact that the ellipse fitting method misses some of bulge-dominant barred galaxies and the Fourier analysis finds strongly barred galaxies.

\end{enumerate}

We thank the anonymous referee for comments that improved this paper. We acknowledge the support for YHL and MGP by the National Research Foundation of Korea to the Center for Galaxy Evolution Research (No. 2017R1A5A1070354). 

{}

\end{document}